\documentclass[a4paper,11pt]{article}
\usepackage{jcappub, bm, color} 
\usepackage{amssymb,amsfonts,slashed,amsthm,amsmath,graphicx, soul, empheq}
\usepackage[caption=false]{subfig}
\begin{document}

\newcommand{\vev}[1]{ \left\langle {#1} \right\rangle }
\newcommand{\bra}[1]{ \langle {#1} | }
\newcommand{\ket}[1]{ | {#1} \rangle }
\newcommand{\eV}{ \ {\rm eV} }
\newcommand{\KeV}{ \ {\rm keV} }
\newcommand{\MeV}{\  {\rm MeV} }
\newcommand{\GeV}{\  {\rm GeV} }
\newcommand{\TeV}{\  {\rm TeV} }
\newcommand{\1}{\mbox{1}\hspace{-0.25em}\mbox{l}}
\newcommand{\Red}[1]{{\color{red} {#1}}}

\newcommand{\lmk}{\left(}  
\newcommand{\rmk}{\right)}
\newcommand{\lkk}{\left[}  
\newcommand{\rkk}{\right]}
\newcommand{\lhk}{\left \{ }  
\newcommand{\rhk}{\right \} }
\newcommand{\del}{\partial}  
\newcommand{\la}{\left\langle} 
\newcommand{\ra}{\right\rangle}
\newcommand{\half}{\frac{1}{2}}

\newcommand{\bea}{\begin{array}}
\newcommand{\eea}{\end{array}}
\newcommand{\beq}{\begin{eqnarray}}
\newcommand{\eeq}{\end{eqnarray}}
\newcommand{\eq}[1]{Eq.~(\ref{#1})}

\newcommand{\dd}{\mathrm{d}}
\newcommand{\Mpl}{M_{\rm Pl}}
\newcommand{\mg}{m_{3/2}}
\newcommand{\abs}[1]{\left\vert {#1} \right\vert}
\newcommand{\mphi}{m_{\phi}}
\newcommand{\Hz}{\ {\rm Hz}}
\newcommand{\for}{\quad \text{for }}
\newcommand{\Min}{\text{Min}}
\newcommand{\Max}{\text{Max}}
\newcommand{\Kahler}{K\"{a}hler }
\newcommand{\cphi}{\varphi}
\newcommand{\Tr}{\text{Tr}}
\newcommand{\diag}{{\rm diag}}

\newcommand{\SUf}{SU(3)_{\rm f}}
\newcommand{\Upq}{U(1)_{\rm PQ}}
\newcommand{\Zpq}{Z^{\rm PQ}_3}
\newcommand{\Cpq}{C_{\rm PQ}}
\newcommand{\ubar}{u^c}
\newcommand{\dbar}{d^c}
\newcommand{\ebar}{e^c}
\newcommand{\nubar}{\nu^c}
\newcommand{\Ndw}{N_{\rm DW}}
\newcommand{\Fpq}{F_{\rm PQ}}
\newcommand{\fpq}{v_{\rm PQ}}
\newcommand{\Br}{{\rm Br}}
\newcommand{\Lag}{\mathcal{L}}
\newcommand{\Lqcd}{\Lambda_{\rm QCD}}

\newcommand{\ji}{j_{\rm inf}} 
\newcommand{\jb}{j_{B-L}} 
\newcommand{\M}{M} 
\newcommand{\im}{{\rm Im} }
\newcommand{\re}{{\rm Re} }

\def\lrf#1#2{ \left(\frac{#1}{#2}\right)}
\def\lrfp#1#2#3{ \left(\frac{#1}{#2} \right)^{#3}}
\def\lrp#1#2{\left( #1 \right)^{#2}}
\def\REF#1{Ref.~\cite{#1}}
\def\SEC#1{Sec.~\ref{#1}}
\def\FIG#1{Fig.~\ref{#1}}
\def\EQ#1{Eq.~(\ref{#1})}
\def\EQS#1{Eqs.~(\ref{#1})}
\def\TEV#1{10^{#1}{\rm\,TeV}}
\def\GEV#1{10^{#1}{\rm\,GeV}}
\def\MEV#1{10^{#1}{\rm\,MeV}}
\def\KEV#1{10^{#1}{\rm\,keV}}
\def\blue#1{\textcolor{blue}{#1}}
\def\red#1{\textcolor{blue}{#1}}

\newcommand{\eff}{\Delta N_{\rm eff}}
\newcommand{\neff}{\Delta N_{\rm eff}}
\newcommand{\cc}{\Omega_\Lambda}
\newcommand{\Mpc}{\ {\rm Mpc}}
\newcommand{\Msolar}{M_\odot}

\def\my#1{\textcolor{blue}{#1}}
\def\MY#1{\textcolor{blue}{[{\bf MY:} #1}]}

%%%%%%%%%%%%%%%%%%%%%%%%%%%%%%%%%%%%%%%%%%%%%%%%%%%%%%%%%%%%%%%

%######################
\begin{flushright}
TU-1257\\
KEK-QUP-2025-0006
\end{flushright}
%######################

\title{
Decay rate of PQ-ball
}

\author{Jin Kobayashi$^{1}$, 
Kazunori Nakayama$^{1,2}$,
Masaki Yamada$^{1}$}
\affiliation{$^{1}$Department of Physics, Tohoku University, Sendai, Miyagi 980-8578, Japan}
\affiliation{$^{2}$International Center for Quantum-field Measurement Systems for Studies of the Universe and Particles (QUP), KEK, Tsukuba, Ibaraki 305-0801, Japan}

\abstract{
Q-balls are non-topological solitons that arise in theories with a complex scalar field possessing a conserved global U(1) charge. Their stability is ensured by this charge, making them potentially significant in cosmology. In this paper, we investigate Q-ball-like objects in scenarios where the scalar field acquires a finite vacuum expectation value, spontaneously breaking the global U(1) symmetry. A well-motivated example is the Peccei-Quinn field, where the U(1) symmetry is identified as the Peccei-Quinn symmetry, and hence we refer to such objects as PQ-balls. We first discuss the existence of stable PQ-ball solutions in a finite-density plasma and argue that they become unstable in vacuum. Using detailed numerical simulations under spherical symmetry, we confirm their formation, compute their decay rate, and derive an analytical formula for it. Our results have important implications for axion cosmology, particularly in the context of the kinetic misalignment mechanism.
}

\emailAdd{jin.kobayashi.q6@tohoku.ac.jp}
\emailAdd{kazunori.nakayama.d3@tohoku.ac.jp}
\emailAdd{m.yamada@tohoku.ac.jp}

\maketitle
\flushbottom

%%%%%%%%%%%%%%%%%%%%%%%%%%%%%%%%%%%%%%%%%%%%%%%%%%%
\section{Introduction}
%%%%%%%%%%%%%%%%%%%%%%%%%%%%%%%%%%%%%%%%%%%%%%%%%%%

Complex scalar fields play crucial roles in particle physics and cosmology, not only within the Standard Model (SM) but also in its extensions.
For example, they are essential for realizing spontaneous symmetry breaking and for the formation of solitons.
Depending on their potential, they may also serve as the inflaton if their energy dominates the early Universe, or as dark matter candidates, such as when they undergo coherent oscillations in a quadratic potential.

Depending on the shape of the potential, scalar fields can form localized configurations known as solitons.
When their stability is guaranteed by topological considerations, they are referred to as topological defects.
In addition to these, non-topological solitons~\cite{Friedberg:1976me,Lee:1991ax} can also exist, such as Q-balls~\cite{Coleman:1985ki}, which may arise when a complex scalar field possesses a conserved U(1) charge.
Q-balls are localized due to attractive self-interactions and are stabilized by the conservation of the U(1) charge.
In cosmology, such objects can play important roles, potentially altering the thermal history of the Universe through their formation and subsequent decay (see Refs.~\cite{Kusenko:1997si,Enqvist:1997si,Enqvist:1998en,Kasuya:1999wu,Kasuya:2000wx,Kasuya:2000sc,Kasuya:2001hg,Kamada:2012bk,Harigaya:2014tla} in the context for Affleck-Dine baryogenesis).

From a theoretical perspective, it is an intriguing question whether Q-balls can exist in models where the U(1) symmetry is spontaneously broken in the vacuum.
In this paper, we demonstrate that Q-ball-like solutions can exist and remain stable in a finite-density plasma.
Since a particularly interesting example of such a U(1) symmetry is the U(1)$_{\rm PQ}$ symmetry, we refer to these objects as PQ-balls in this paper, as they carry a U(1)$_{\rm PQ}$ charge.
However, unlike conventional Q-balls, these configurations are expected to be unstable in vacuum due to the structure of the potential responsible for the spontaneous symmetry breaking.
We numerically compute the decay rate of this new type of Q-ball in vacuum and discuss its theoretical implications.% 
\footnote{ Note that the decay process considered in this paper differs from the perturbative decay of scalar fields into other particles, as discussed in Refs.~\cite{Cohen:1986ct,Kawasaki:2012gk}.
A PQ-ball can decay even in the absence of interactions between the complex scalar field and other fields. 
}
We also numerically confirm the formation of such Q-balls from small perturbations, assuming spherical symmetry.%
\footnote{
See also the Note Added at the end of this paper for a related study in a two-field model that appeared on the same day~\cite{accompanypaper}.
The configurations discussed therein differ from ours and are not identified as PQ-balls.
}

We also discuss the possibility that such objects may form in a well-motivated cosmological scenario and play an essential role in the thermal history of the Universe.
A primarily interesting example for spontaneously broken U(1) symmetry is U(1)$_{\rm PQ}$ symmetry, which is introduced and is assumed to be spontaneously broken to solve the strong CP problem~\cite{Peccei:1977hh,Peccei:1977ur}. 
Associated pseudo Nambu-Goldstone boson, axion, 
is a candidate of dark matter in the universe~\cite{Weinberg:1977ma,Wilczek:1977pj}.
A conventional production mechanism of axion dark matter is the so-called misalignment mechanism~\cite{Preskill:1982cy,Abbott:1982af,Dine:1982ah}.
However, the cosmological dynamics of the axion and the PQ-symmetry-breaking field can be much more involved, and several other production mechanisms have also been investigated~\cite{Kawasaki:2013ae,DiLuzio:2020wdo}.

The kinetic misalignment mechanism~\cite{Co:2019jts,Chang:2019tvx} has recently been proposed as a new axion production mechanism.
In this scenario, the radial component of the complex scalar field $\Phi$, which breaks the PQ symmetry, plays a central role.
It is assumed that the U(1)$_{\rm PQ}$ charge is generated through the dynamics of $\Phi$ in the early Universe, via a mechanism analogous to the Affleck-Dine mechanism~\cite{Affleck:1984fy,Dine:1995kz}.
As is well known in the context of Affleck-Dine baryogenesis, such charge densities can become unstable depending on the shape of the scalar potential, potentially leading to the formation of (P)Q-balls.
Similarly, if (P)Q-balls form, the subsequent cosmological evolution changes accordingly.
In particular, the dissipation and thermalization of the saxion field may be affected by the formation of PQ-balls.\footnote{
Refs.~\cite{Fonseca:2019ypl,Eroncel:2022vjg,Eroncel:2022efc} considered the axion fragmentation in the kinetic misalignment mechanism. While they focused on the axion dynamics, our findings are essentially related to the radial dynamics of the PQ field.
}

This paper is organized as follows.
In Sec.~\ref{sec:PQ} we overview the PQ-ball solution in comparison with the conventional Q-ball solution.
In Sec.~\ref{sec:num_sim} we numerically simulate the time evolution of the PQ-ball solution to find the PQ-ball decay rate or lifetime.
We also give a theoretical interpretaion of the result.
In Sec.~\ref{sec:app} some cosmological applications of our results are presented.
Sec.~\ref{sec:discussions} is devoted to conclusions and discussion.

\section{PQ-ball solution}
\label{sec:PQ}

\subsection{Overview of PQ-ball formation scenario}

Although the primary focus of this paper is on the properties of PQ-balls after their formation, we briefly discuss a cosmological scenario for their formation in the context of the kinetic misalignment mechanism.

For the kinetic misalignment mechanism to be effective, there must be a mechanism in the early universe that induces the rotation of the PQ symmetry-breaking field 
$\Phi$ in the phase direction. Such a mechanism is analogous to the Affleck-Dine mechanism, which generates baryon number in SUSY theories~\cite{Affleck:1984fy,Dine:1995kz}.
Thus below we briefly recall the basic idea of the Affleck-Dine mechanism.

In SUSY theories, some combinations of scalar fields can have a very flat potential~\cite{Dine:1995kz,Gherghetta:1995dv}. 
Considering the dynamics along such flat directions, it is possible for the field
$\Phi$ to acquire a large vacuum expectation value (VEV) in the early Universe~\cite{Dine:1995uk}.
As the universe expands and the Hubble parameter decreases, the effect of the mass term $m_s^2 \abs{\Phi}^2$ leads to oscillations around the origin. 
At this stage, the higher-dimensional potential terms introduce a gradient in the phase direction, causing the complex scalar field $\Phi$ to rotate in phase space. 
This rotation corresponds to the generation of a U(1) charge. However, due to redshift, the higher-dimensional potential terms quickly become negligible, leading to the termination of U(1) charge generation.
In the subsequent evolution of the universe, the U(1) symmetry of 
$\Phi$ is restored, and the generated U(1) charge remains conserved. 
This is the basic scenario of the Affleck-Dine mechanism~\cite{Affleck:1984fy,Dine:1995kz}.

In the above discussion, $m_s$ represents the effective mass of 
$\Phi$, which generally depends on the field itself. Depending on the functional form of the effective mass, fluctuations in the condensate can grow, leading to the formation of Q-balls, a type of non-topological soliton~\cite{Kusenko:1997ad,Kusenko:1997zq}.
The formation of Q-balls has significant impacts on the cosmological consequences, such as the final baryon number or dark matter abundance~\cite{Kusenko:1997si,Enqvist:1997si,Enqvist:1998en,Kasuya:1999wu,Kasuya:2000wx,Kasuya:2000sc,Kasuya:2001hg,Kamada:2012bk,Harigaya:2014tla}.

In the Affleck-Dine mechanism, the U(1) charge corresponds to baryon number and lepton number. However, in a similar theoretical framework, the same dynamics can generate other U(1) charges. 
In the kinetic misalignment mechanism, a similar dynamical process generates the $\text{U(1)}_\text{PQ}$ charge. In this context, 
this radial component is regarded as the saxion in the context of supersymmetry (SUSY)~\cite{Rajagopal:1990yx,Kim:1991van,Lyth:1993zw}. 
Specifically, the field 
$\Phi$ is composed of both the saxion and axion as components of a complex scalar field. 
It is known that a U(1) symmetry generally ensures a flat direction in the scalar potential in SUSY~\cite{Kugo:1983ma}. 
Thus, it is plausible that the dynamics of the saxion resemble those of the Affleck-Dine field, potentially leading to the formation of localized configurations, which we refer to as PQ-balls.
In fact, linear perturbation theory suggests that a homogeneously rotating complex scalar field is unstable and tends to develop a localized configuration, as discussed in Appendix~\ref{sec:linear}.
In the following, we briefly review the standard Q-ball solution and then explain the case with spontaneously broken PQ symmetry.

\subsection{Q-ball}

S.~Coleman discovered a spatially localized solution in a system with a global U(1) symmetry and named this solution the Q-ball~\cite{Coleman:1985ki}. The letter ``Q'' emphasizes that the solution is stabilized by the conservation of U(1) charge:
\begin{equation}
Q = i\int d^{3}x (\Phi\dot{\Phi}^{*}-\Phi^{*}\dot{\Phi}).
\label{Q_charge}
\end{equation}

In the standard treatment of Q-balls, it is typically assumed that the Q-ball is localized within a finite region while the volume of the system is taken to be infinite. For our purposes, however, we consider a finite spherical volume of radius $L$, which we assume to be sufficiently large. This parameter serves as an infrared (IR) cutoff. Although a localized Q-ball is formally defined in the limit $L \to \infty$, in practice, the solution is unaffected as long as $L$ is much larger than the typical size of the Q-ball. In a realistic setting, 
$L$ corresponds to the typical distance between Q-balls in the Universe.
This IR cutoff is also necessary when considering the case of a finite-density plasma, as discussed below. 
Hereafter, we implicitly assume that all volume integrals are performed over the finite spherical region of sufficiently large radius $L$.

Our goal now is to minimize the energy while keeping the charge constant. We apply the method of Lagrange multipliers, introducing a Lagrange multiplier $\omega$ to fix $Q$. The function to be minimized is
\begin{equation}
E_{\omega} = E + \omega\left[ Q - \frac{i}{2}\int d^{3}x (\varphi\dot{\varphi}^{*}-\varphi^{*}\dot{\varphi}) \right],
\end{equation}
where $E$ is the total enrgy of the system and we have defined $\varphi \equiv \sqrt{2}\Phi$ for convenience.
Substituting the expression for $E$, this can be rewritten as
\begin{equation}
E_{\omega} = \frac{1}{2}\int d^{3}x|\dot{\varphi}-i\omega\varphi|^{2} + \int d^{3}x\left[ \frac{1}{2}|\partial_{i}\varphi|^{2} + V_{\omega}(|\varphi|) \right] + \omega Q,
\label{eq:E_omega}
\end{equation}
where 
\begin{equation}
V_{\omega}(|\varphi|) = V(|\varphi|) - \frac{1}{2}\omega^{2}|\varphi|^{2}.
\label{eq:V_omega}
\end{equation}

By imposing time dependence on the phase of the field,
\begin{equation}
\varphi(\vec{x},t) = e^{i\omega t}\phi(\vec{x}),
\label{eq:min_t}
\end{equation}
the first term in Eq.~(\ref{eq:E_omega}) vanishes. Here, $\phi(\vec{x})$ is a real scalar field that depends only on the spatial coordinates. While it is possible to consider a spatially dependent phase, this would contribute positively to the spatial derivative term in Eq.~(\ref{eq:E_omega}), preventing minimization. Additionally, removing any physically meaningless constant phase, we can take $\phi(\vec{x})$ to be real.

The remaining part of the functional to be minimized is
\begin{equation}
E_{\omega} = \int d^{3}x\left[ \frac{1}{2}|\partial_{i}\phi(\vec{x})|^{2} + V_{\omega}(\phi(\vec{x})) \right] + \omega Q.\label{eq:edens1}
\end{equation}
By applying the variational principle, the equation of motion for the bounce solution is derived as
\begin{equation}
\frac{d^{2}\phi(r)}{dr^{2}} + \frac{2}{r}\frac{d\phi(r)}{dr} + \left[\omega^{2}\phi(r)-\frac{\partial V(\phi)}{\partial\phi}\right] = 0.
\label{eq:eom}
\end{equation}
This solution $\bar{\phi}_{\omega}(r)$ minimizes the functional and satisfies the following boundary conditions:
\begin{equation}
\left.\frac{d\phi}{dr}\right|_{r\rightarrow0} = 0, \ \ \phi(r = L) = 0. \label{eq:bc1}
\end{equation}
The first condition ensures that the field remains constant near the origin and avoids singularities. The second condition enforces that the field approaches the vacuum state $\phi=0$ at the boundary of the system. 
In the limit of $L \to \infty$, this boundary condition is essential to ensure that the total energy of the system remains finite. 
This is the method to determine the Q-ball profile for the case with a global U(1) symmetry that is not spontaneously broken at vacuum.

As an example, let us consider the following potential for the complex scalar field:
\begin{equation}
    V = m^2 \left[1 + K \log\frac{|\Phi|^2 + v^2}{M^2} \right] |\Phi|^2,
    \label{eq:gravitymed}
\end{equation}
where $K<0$ is a constant and $v$ and $M$ are some mass scales. 
The Q-ball profile is approximated to be given by a Gaussian form~\cite{Enqvist:1998en}: 
\begin{equation}
    \phi(r) = \phi_0 \exp \left(-\frac{r^2}{R^2}\right),~~~~~~~~~~R = \frac{\sqrt{2}}{m\sqrt{|K|}}.
\end{equation}
This is sometimes called gravity-mediation Q-balls because the potential is motivated by gravity-mediated SUSY-breaking models.

\subsection{PQ-ball}
\label{sec:PQQball}

\begin{figure}[t]
\centering
\includegraphics[width=0.7\linewidth]{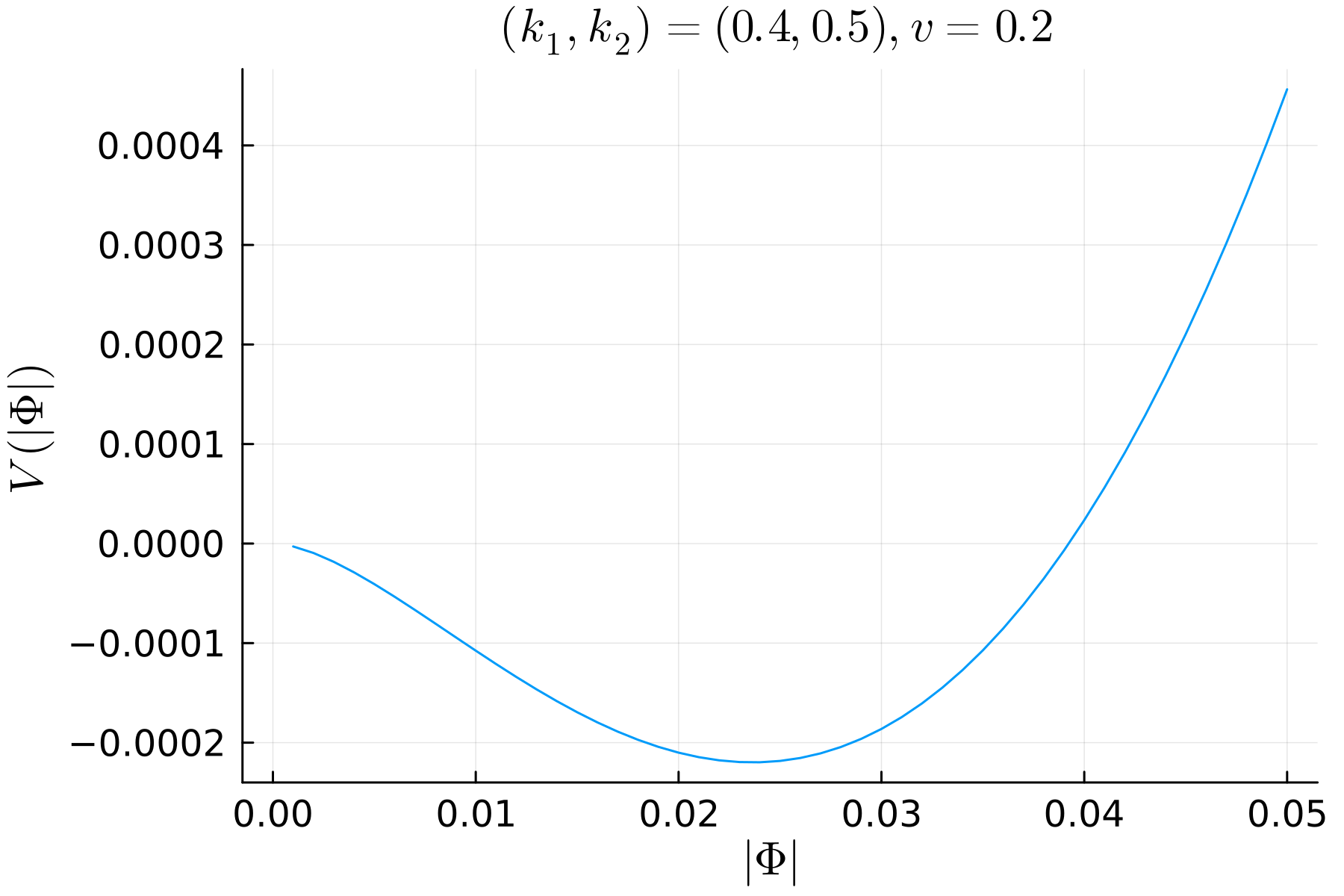}
\caption{Profile of the potential $V$ for $k_{1}=0.4,k_{2}=0.5,v=0.2$ with $m=M=1$. 
This potential has a local minimum away from the origin, enabling spontaneous symmetry breaking.}
\label{fig:Vax1}
\end{figure}

We now consider a Q-ball-like solution in a theory with spontaneous symmetry breaking of U(1) at a vacuum. 
We assume that the potential can be written as
\begin{equation}
	V(\Phi) = m^{2}\left[ 1+k_{1}\log\frac{|\Phi|^{2}}{M^{2}} - k_{2}\log\frac{|\Phi|^{2}+v^{2}}{M^{2}} \right] |\Phi|^{2} ~~~~~~~~(k_{2}>k_{1}>0),
	\label{eq:potential_PQ}
\end{equation}
(see Fig.~\ref{fig:Vax1}). The specific theory and derivation of this potential are discussed in App.~\ref{sec:model}. Here, $v$ is the energy scale where the running mass behavior of 
$\Phi$ changes, and it is  assumed to be smaller than another mass scale $M$. Since this potential takes negative values for small 
$\abs{\Phi}$, it allows spontaneous breaking of PQ symmetry if $\Phi$ has a U(1)$_{\rm PQ}$ charge.\footnote{
    Such a ``radiative stabilization'' of the PQ field has been previously considered in Refs.~\cite{Abe:2001cg,Nakamura:2008ey,Moroi:2014mqa}.
}
The VEV of $\Phi$ at the potential minimum, denoted by $f_a =\sqrt{2}\left< |\Phi| \right>$, is the PQ scale and it gives the decay constant of the axion.
On the other hand, for sufficiently large $\abs{\Phi}$ relative to $v$, the potential simplifies to
\begin{equation}
	V(\Phi) \simeq m^{2}\left[ 1+(k_{1}-k_{2})\log\frac{|\Phi|^{2}}{M^{2}} \right] |\Phi|^{2},
\end{equation}
which resembles the gravity-mediation-type potential when 
$k_{1}-k_{2}=K<0$. Thus, we expect that this potential approximately leads to solutions similar to those found in the gravity-mediation scenario.

Assuming spherical symmetry, the charge retains the same form as before. The energy functional is given by
\begin{equation}
	E = 4\pi\int^{\infty}_{0}\left[\left| \frac{\partial\Phi}{\partial t} \right|^{2} + |\nabla\Phi|^{2}
	+ m^{2}\left( 1+k_{1}\log\frac{|\Phi|^{2}}{M^{2}}-k_{2}\log\frac{|\Phi|^{2} + v^{2}}{M^{2}}\right)|\Phi|^{2}\right]r^{2}dr.
\end{equation}
The Euler-Lagrange equation for $\Phi$ is given by
\begin{multline}
	0=\frac{\partial^{2}\Phi}{\partial t^{2}} - \frac{\partial^{2}\Phi}{\partial r^{2}} - \frac{2}{r}\frac{\partial\Phi}{\partial r} \\
	+ m^{2}\left[ 1+k_{1}\left( 1 + \log\frac{|\Phi|^{2}}{M^{2}} \right)-k_{2}\left( \frac{|\Phi|^{2}}{|\Phi|^{2} + v^{2}} + \log\frac{|\Phi|^{2} + v^{2}}{M^{2}} \right) \right] \Phi.
\label{eq:EL2}
\end{multline}

If the time dependence is confined to the phase,
$\Phi(r,t) = e^{i\omega t}\Phi_{s}(r)$, 
the stationary solution 
$\Phi_{s}(r)$ satisfies the equation:
\begin{multline}
	\frac{\partial^{2}\Phi_{s}}{\partial r^{2}} + \frac{2}{r}\frac{\partial\Phi_{s}}{\partial r} + \omega^{2}\Phi_{s} \\
	= m^{2}\left[ 1+k_{1}\left( 1 + \log\frac{\Phi_{s}^{2}}{M^{2}} \right)-k_{2}\left( \frac{\Phi_{s}^{2}}{\Phi_{s}^{2} + v^{2}} + \log\frac{\Phi_{s}^{2} + v^{2}}{M^{2}} \right) \right]\Phi_{s}.
\end{multline}
From this, the effective potential $U_{\omega}$
 in the mechanical analogy is given by
\begin{equation}
	U_{\omega} = \frac{1}{2}\omega^{2}\Phi_{s}^{2}-\frac{1}{2}m^{2}\left[ 1+k_{1}\log\frac{\Phi_{s}^{2}}{M^{2}}-k_{2}\log\frac{\Phi_{s}^{2}+v^{2}}{M^{2}} \right]\Phi_{s}^{2}.
\label{eq:axU}
\end{equation}

\begin{figure}
    \centering
\includegraphics[width=0.7\linewidth]{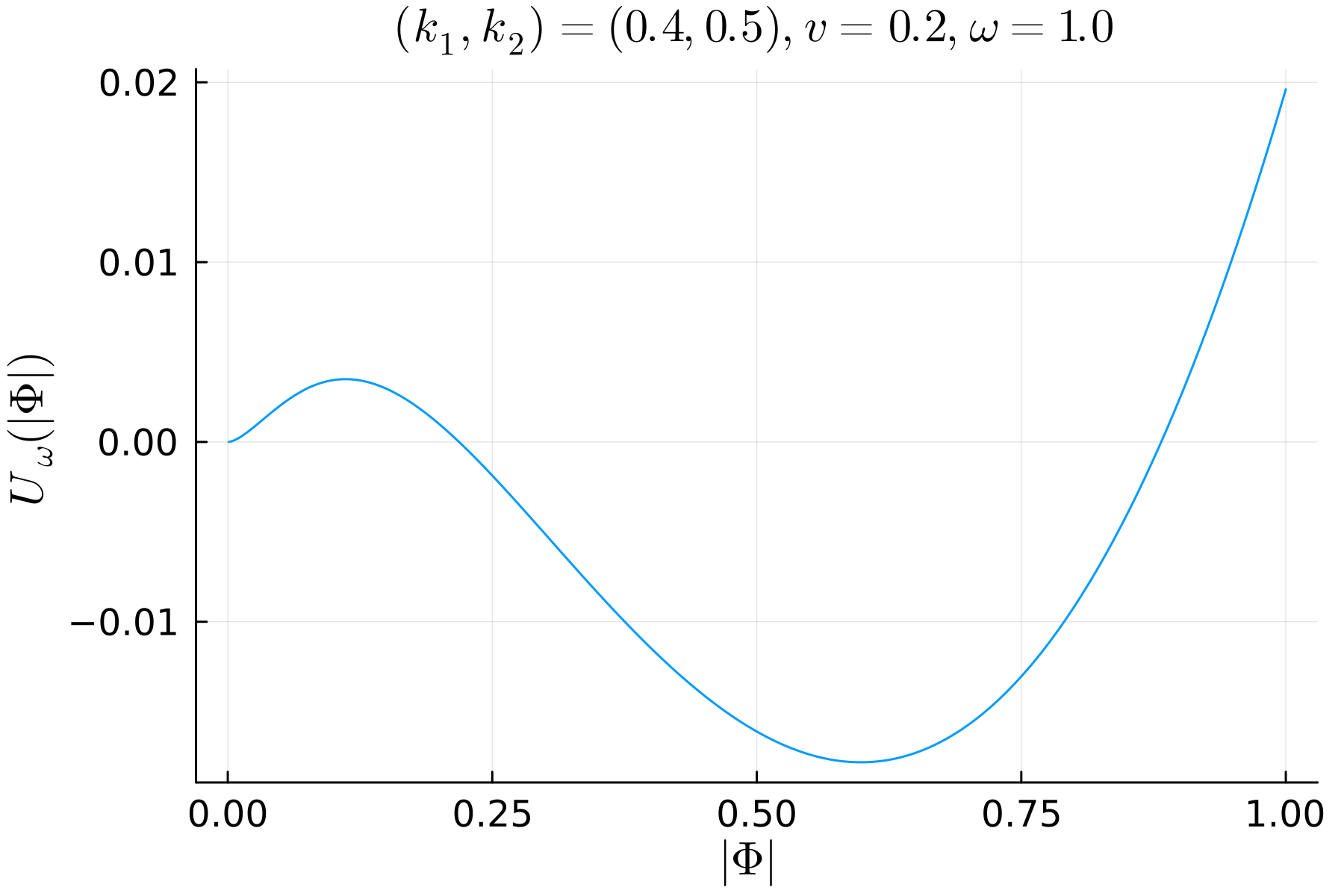}
    \caption{$U_\omega$ for $k_{1}=0.4,k_{2}=0.5,v=0.2,\omega=1$. 
    }    
    \label{fig:Uax}
\end{figure}

As shown in Fig.~\ref{fig:Uax}, a local maximum appears before reaching the origin ($\Phi_{s}=0$). 
Due to this maximum, solutions that asymptotically approach 
$\Phi_{s}=0$ at 
$r = L \rightarrow\infty$ do not exist. Instead, there exists a solution where the asymptotic field value is not zero but rather a positive finite value 
$\Phi_{s\infty}$. This suggests that the solution corresponds to a Q-ball with a surrounding charge density, which cannot exist stably in vacuum.\footnote{Such solutions are sometimes referred to as Q-bulges \cite{Nugaev:2016wyt,Nugaev:2019vru}. They focused on the existence of stable configurations in a finite-density plasma, whereas our work concerns the longevity of similar configurations in vacuum.} 
We refer to such metastable configurations in a model with spontaneously broken U(1) symmetry as PQ-balls, as they are primarily motivated by a PQ model.
Importantly, $\Phi_{s\infty}$ is not necessarily equal to the vacuum expectation value 
$f_{a}$ of the PQ field, as the extrema of $U_{\omega}$ and the original potential 
$V$ do not coincide: in general we have $\Phi_{s\infty} > f_a$.
Note that the total charge $Q$ (\ref{Q_charge}), as well as the energy, is formally divergent in the limit of $L \to \infty$ because of finite $\Phi_{s\infty}$. 
However, for a large but finite system size $L$, the total charge remains finite, and the problem of finding the lowest-energy state at fixed charge is well-defined. One can always take the large-$L$ limit, in which the equations of motion and the Q-ball solution remain essentially unchanged.%
\footnote{
Instead, one may redefine $Q$ as the difference from the background charge~\cite{Nugaev:2016wyt,Nugaev:2019vru}.
    Note that $\Phi(r,t)=\Phi_{s\infty} e^{i\omega t}$ is also a trivial solution to the equation of motion.
}

\begin{figure}
\centering
\includegraphics[width=0.7\linewidth]{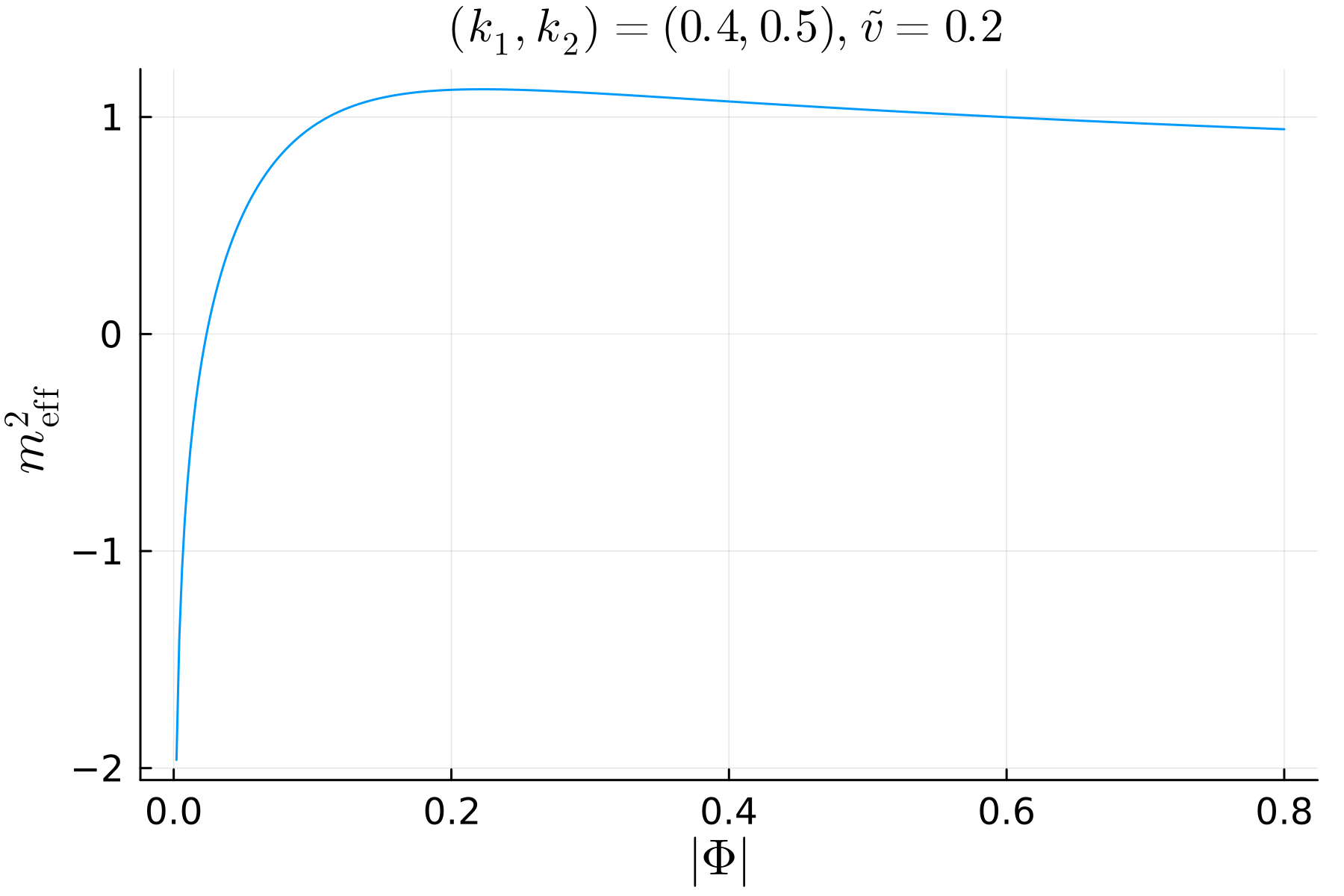}
\includegraphics[width=0.7\linewidth]{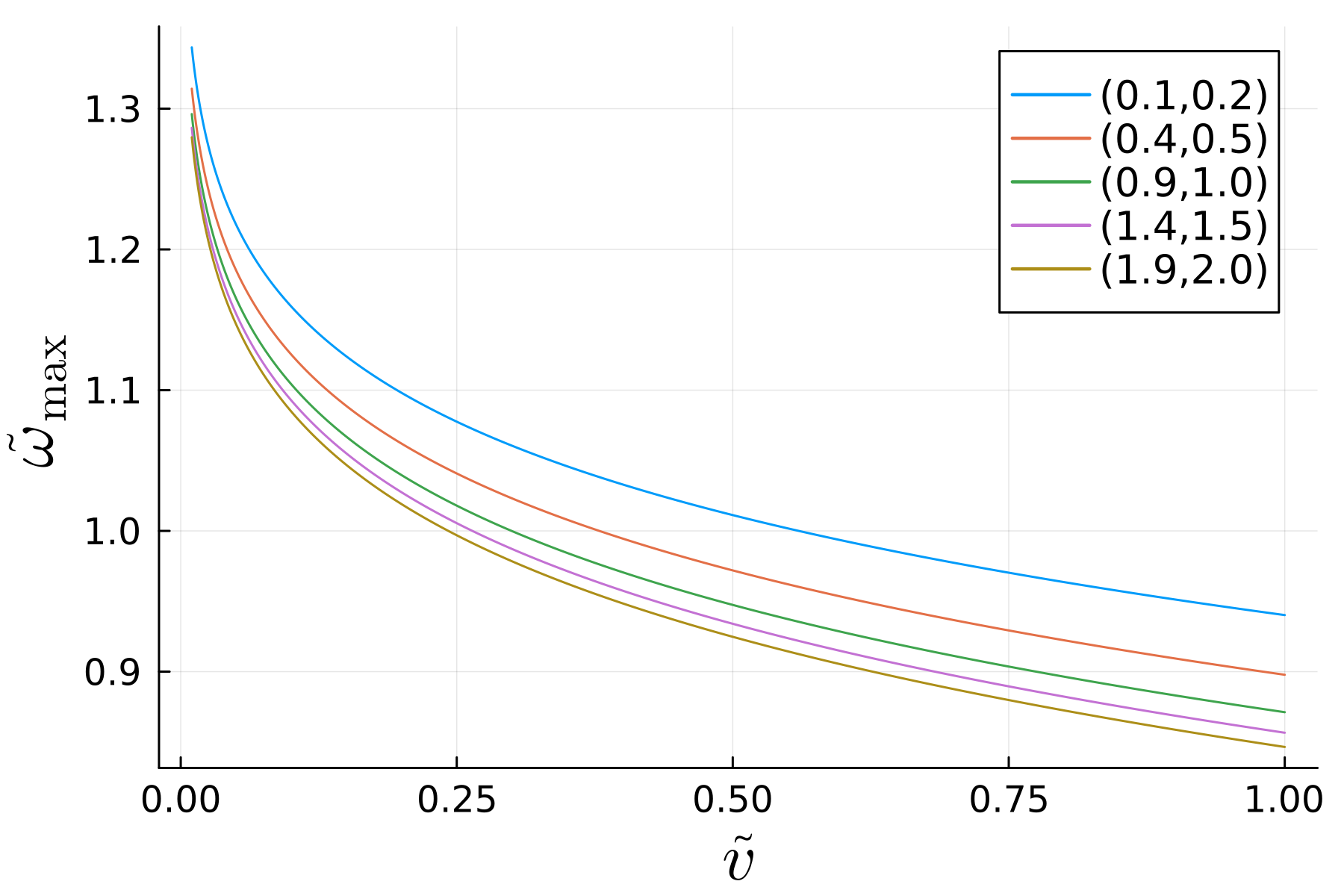}
\caption{
Top: Dependence of $m_\text{eff}^{2}$ against $|\Phi|$ for
$(k_{1},k_{2})=(0.4,0.5),\tilde{v}\equiv v/m=0.2$. Near the origin, 
$m_\text{eff}^{2}$ is negative and hence tachyonic, but it reaches a local maximum at a certain 
$|\Phi|$ before decreasing monotonically. If $\omega^{2}$ is smaller than this maximum value, 
$U_{\omega}$ develops a local maximum, allowing for the existence of a Q-ball solution.
Bottom: Dependence of $\omega_\text{max}$ against three parameters $\tilde v, k_1, k_2$. The legend in the upper right corner indicates the values of $(k_{1},k_{2})$.
}
\label{fig:effmass2}
\end{figure}

For spherically symmetric condensate solutions to exist, 
$U_{\omega}$ must exhibit a local maximum near the origin. This imposes an upper bound on 
$\omega$, as solutions disappear when 
$\omega$ exceeds this maximum. The top panel of Fig.~\ref{fig:effmass2} illustrates the behavior of 
$m_\text{eff}^{2}(\Phi) \equiv V / |\Phi|^2$, showing that a local maximum exists, beyond which it decreases monotonically.
Thus we must have\footnote{
    Literally the solution always exists for any small $\omega$. However, $\Phi_{s0}$ becomes larger for smaller $\omega$ and we should take account of higher order corrections to the scalar potential. We will consider $\omega$ not very far from $\omega_{\rm max}$ in the following discussion.
}
\begin{equation}
	\omega < \omega_\text{max},
\end{equation}
where $\omega_\text{max}$ depends on $(k_{1},k_{2},v)$. 
Numerical results confirm that smaller $v$ and $(k_1, k_2)$ lead to larger $\omega_\text{max}$.

In summary, PQ-balls resemble gravity-mediation Q-balls when
$\Phi_{s0}\gg v$, but they asymptote to a finite value $\Phi_{s}\rightarrow\Phi_{s\infty}>0$ at large distance. Since the charge density is 
$\omega\Phi_{s}^{2}$, this suggests that PQ-balls are unstable in vacuum. 
Investigating their decay process and determining the decay rate is the focus of the next section.

\section{Numerical simulation of a PQ-ball}
\label{sec:num_sim}

In the PQ model, similar arguments to those of the Affleck-Dine field suggest that the field $\Phi$ may form Q-ball-like objects after the onset of rotation in the phase space, which we call PQ-balls (see Appendix~\ref{sec:linear} for a linear perturbaiton theory).
However, it is important to note that PQ-balls exist as stationary solutions only for a finite charge distribution. If the homogeneous condensate surrounding the PQ-ball disappears due to cosmic expansion, the PQ-ball becomes unstable against decay into individual particles, leading to a reduction of the PQ-ball’s U(1) charge. 
In this section, we numerically simulate the decay of PQ-balls and determine the decay rate.

\subsection{Equation of motion}
For numerical purposes, we define dimension-free parameters: $\chi=\Phi/M$, $\rho=mr,\tau=mt$.
Assuming spherical symmetry, the Euler-Lagrange equation in terms of these dimensionless quantities is rewritten as follows:
\begin{equation}
\frac{\partial^{2}\chi}{\partial\tau^{2}} - \frac{\partial^{2}\chi}{\partial\rho^{2}}
- \frac{2}{\rho}\frac{\partial\chi}{\partial\rho} +  \left[ 1+k_{1}\left( 1 + \log|\chi|^{2} \right)
-k_{2}\left( \frac{|\chi|^{2}}{|\chi|^{2} + \tilde{v}^{2}} + \log( |\chi|^{2} + \tilde{v}^{2} ) \right) \right] \chi = 0,
\label{eq:EL3}
\end{equation}
where $\tilde{v}$ is the dimensionless vacuum expectation value, defined as $\tilde{v}=v/M$. Similarly, the charge and energy are expressed in terms of dimensionless parameters as follows:
\begin{align}
	Q &= 8\pi m^{-2}M^{2}\int^{\infty}_{0}\left(\chi_{1}\frac{\partial\chi_{2}}{\partial\tau}-\frac{\partial\chi_{1}}{\partial\tau}\chi_{2}\right)\rho^{2}d\rho 
	\label{eq:charge_rescale}
    \\
    &\equiv  m^{-2}M^{2} q,
\end{align}
\begin{equation}
	E = 4\pi m^{-1}M^{2}\int^{\infty}_{0}\left[ \left| \frac{\partial\chi}{\partial\rho} \right|^{2} + \left| \frac{\partial\chi}{\partial\tau} \right|^{2} + \left(1+k_{1}\log|\chi|^{2}-k_{2}\log(|\chi|^{2} + \tilde{v}^{2})\right)|\chi|^{2}\right]\rho^{2}d\rho.
\end{equation}
The free parameters in this setup are $k_{1},k_{2},\tilde{v},\tilde{\omega}(\equiv\omega/m)$, giving a total of four independent parameters.

\subsection{Formation of PQ-balls}
\label{formation}

For the sake of completeness, we first examine the formation of PQ-balls in certain regions of parameter space before discussing the details of their decay.

Since the primary motivation for PQ-ball formation during the thermal history of the Universe is the kinetic misalignment mechanism, we assume that the PQ-breaking scalar $\Phi$ initially undergoes coherent rotation in phase space. Therefore, in this subsection, the initial conditions at $\tau=0$ are taken to be
\begin{equation}
\chi|_{\tau=0} = \chi_{c}, \ \ \ \ \dot{\chi}|_{\tau=0} = i\tilde{\omega}\chi_{c}
\end{equation}
where $\dot{\chi}$ represents the derivative with respect to $\tau$, and $\chi_{c}$ is a constant parameter. 
This setup represents a homogeneous complex scalar field rotating in phase space. The time evolution of  $\Phi$ is governed by Eq.~(\ref{eq:EL3}). 
We apply the absorbing boundary conditions at a large radius, as described in Appendix~\ref{sec:boundary}, although the formation of the PQ-ball is also confirmed under other boundary conditions.

\begin{figure}
\centering
\includegraphics[width=0.4\linewidth]{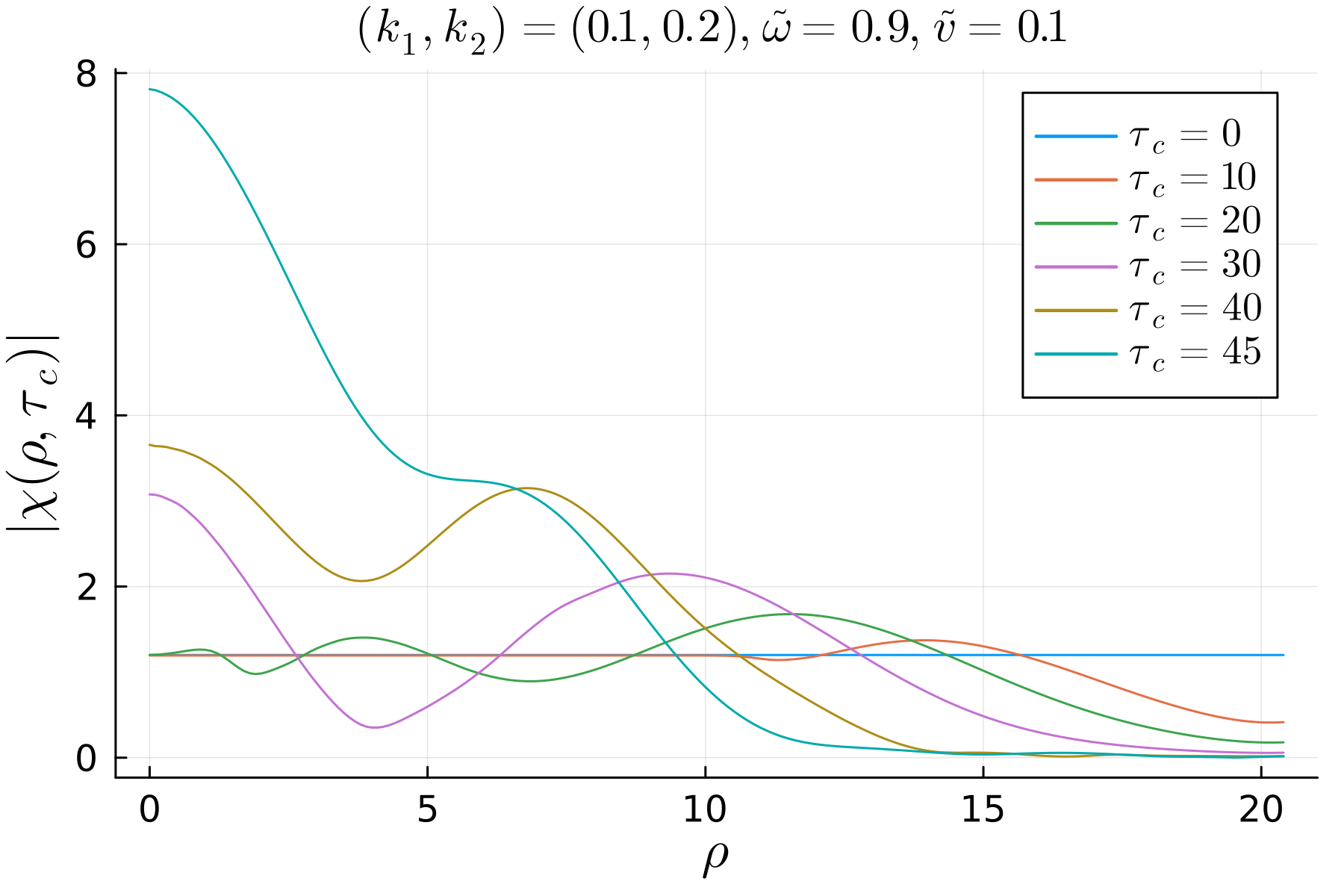}
\hspace{0.3cm}
\includegraphics[width=0.4\linewidth]{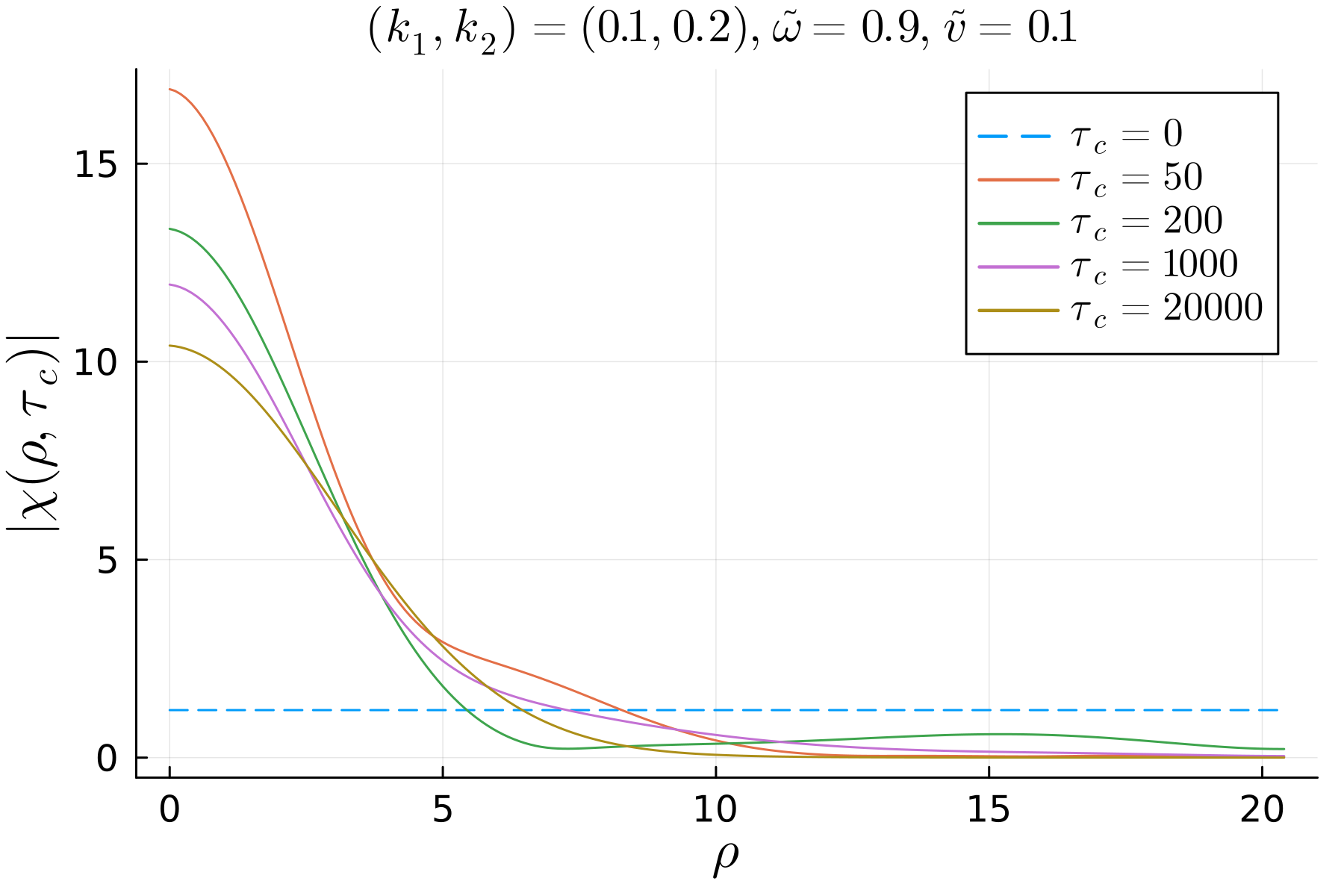}
\vspace{0.6cm}
\\
\includegraphics[width=0.4\linewidth]{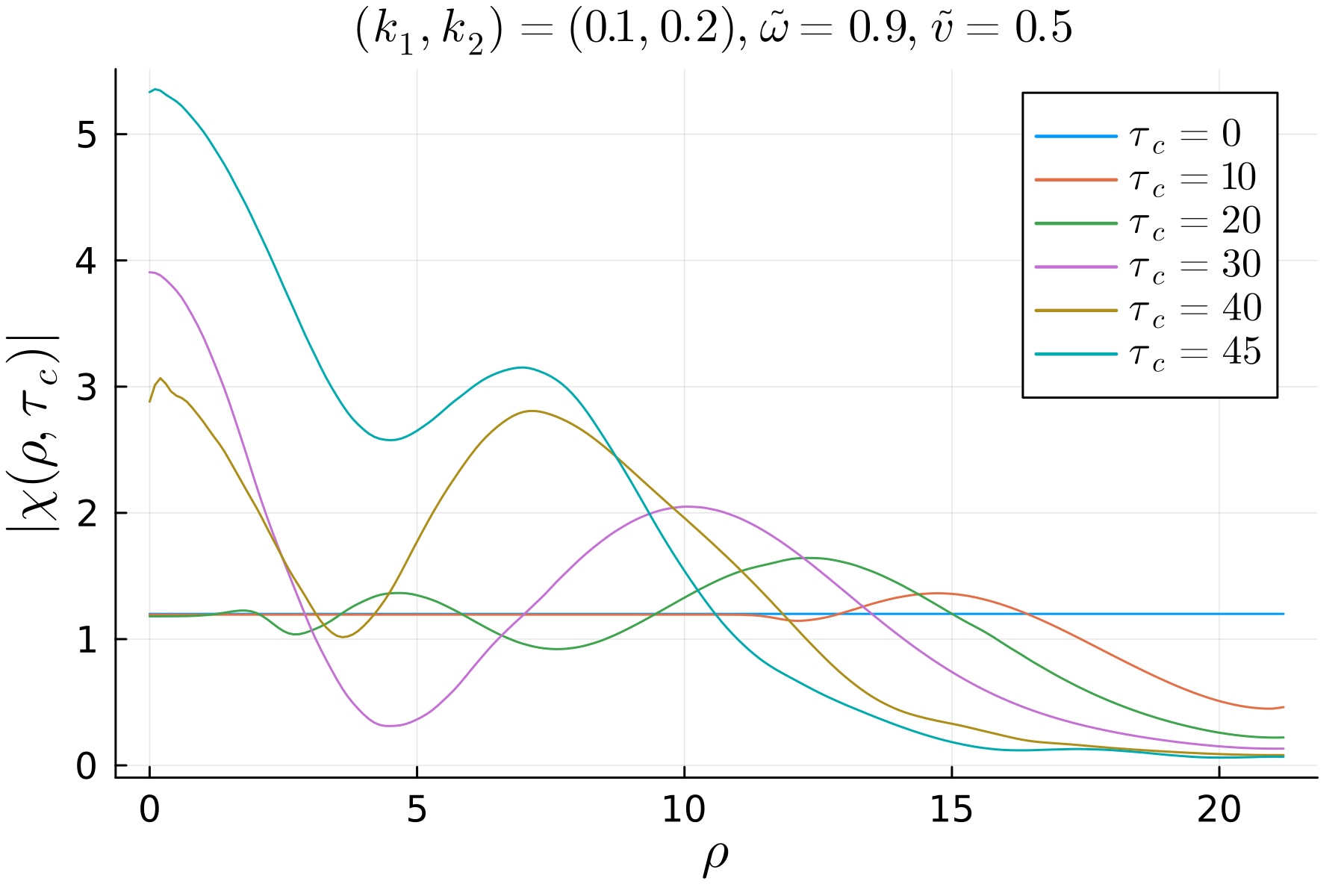}
\hspace{0.3cm}
\includegraphics[width=0.4\linewidth]{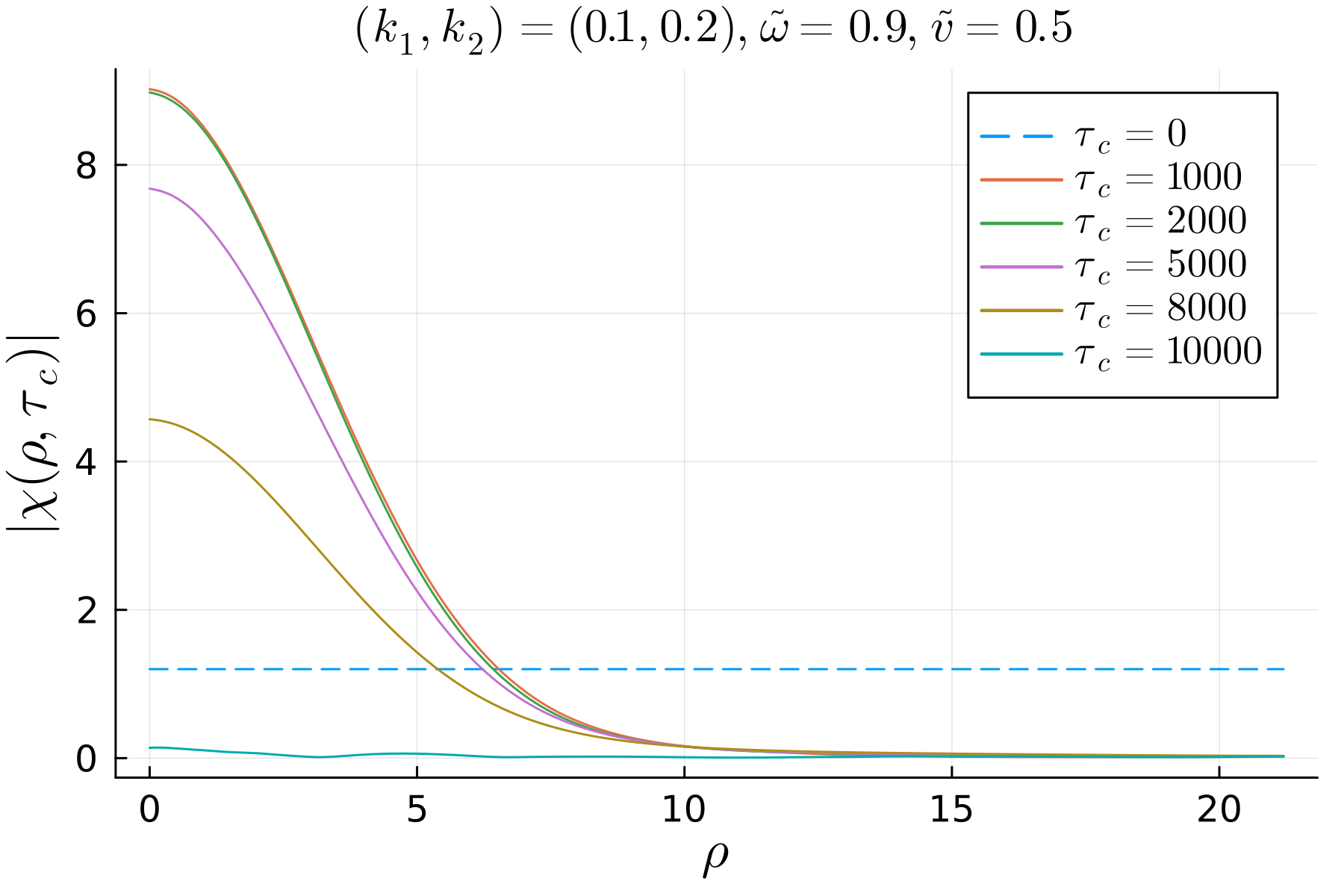}
\caption{
The amplitude $\abs{\chi}$ as a function of $\rho$ at different time steps, where we set $(k_1, k_2) = (0.1, 0.2)$, $\tilde{\omega} = 0.9$, $\chi_{c} = 1.2$, and $\tilde{v} = 0.1$ (top panels) or $\tilde{v} = 0.5$ (bottom panels). 
The dashed lines in the left panels represent the initial amplitude. 
}
\label{fig:production}
\end{figure}

Figures~\ref{fig:production} show $\abs{\chi}$ as a function of $\rho$ at different time steps, where we set $(k_1, k_2) = (0.1, 0.2)$, $\tilde{\omega} = 0.9$, $\chi_{c} = 1.2$, and $\tilde{v} = 0.1$ (top panels) or $\tilde{v} = 0.5$ (bottom panels). 
In the top left panel, we plot the configuration at $\tau = \tau_c$ for $\tau_c = 0, 10, 20, 30, 40, 45$. 
Due to the absorbing boundary condition, the amplitude in the large-$\rho$ region decreases over time, whereas the amplitude near
$\rho = 0$ grows, leading to the formation of a localized configuration. 
The scales of the growth mode and growth rate are consistent with the predictions of linear perturbation theory, as given by \eq{eq:grouthmode} and \eq{eq:grouthrate}, with
$k_2 - k_1 = 0.1$ and $\dot{\Omega}/m = \tilde{\omega} = 0.9$. 
In the top right panel, we plot the configuration at $\tau = \tau_c = 0$ (dashed line), $50, 200, 1000, 20000$. 
Although the PQ-ball amplitude continues to oscillate over this timescale, a stable localized configuration is clearly observed.

The bottom panels illustrate the case with a larger $\tilde{v}$, 
which results in a higher PQ-ball decay rate, as we will discuss shortly. In this case, the simulation captures both the formation and decay of the PQ-ball. The bottom left panel displays the configuration at $\tau = \tau_c = 0, 10, 20, 30, 40, 45$. 
The qualitative similarity to the top left panel supports the expectation that, at the initial stage of PQ-ball formation, the linear perturbation theory is largely insensitive to the specific form of the potential at small $\Phi$. 
The bottom right panel presents the configuration at $\tau = \tau_c = 0$ (dashed line), $1000, 2000, 5000, 8000, 10000$ (solid curves from top to bottom). 
Over this timescale, the configuration remains stable but gradually decreases in amplitude. Eventually, the PQ-ball almost completely disappears at $\tau \sim 10000$. 
This simulation thus demonstrates both the formation and decay of the PQ-ball across different time scales.

We note that the primary objective of this paper is to determine the decay rate of PQ-balls. A detailed analysis of their formation, such as PQ charge, formation time, and comparison with linear perturbation theory, is left for future work.

\subsection{Static solution}
\label{static_solution}
In the following, we focus on determining the decay rate of PQ-balls, assuming they have already formed.
The computational procedure for PQ-ball decay consists of the following steps: First, we calculate the stationary solution, which is a stable configuration within a finite-density environment. This is explained in this subsection.
Then we assume that such an environment disappears due to cosmic expansion or other effects. In vacuum, the stationary solution is expected to become unstable and start to decay. To investigate this decay behavior, we numerically solve equation (\ref{eq:EL3}) under appropriate boundary conditions in Sec.~\ref{sec:evolution}.

As an initial condition, we assume that the complex scalar field has a constant angular velocity $\omega$ in the phase space:
\begin{equation}
	\chi(\rho,\tau)=e^{i\tilde{\omega}\tau}\chi_{s}(\rho),
	\label{eq:static}
\end{equation}
where $\chi_{s}(\rho)$ is the stationary solution. Substituting this field configuration into the Euler-Lagrange equation (\ref{eq:EL3}), the stationary solution $\chi_{s}$ should satisfie the following differential equation with respect to $\rho$: 
\begin{equation}
\frac{\partial^{2}\chi_{s}}{\partial\rho^{2}} + \frac{2}{\rho}\frac{\partial\chi_{s}}{\partial\rho} 
+ \left( \frac{\omega}{m} \right)^{2}\chi_{s} - \left[ 1+k_{1}\left( 1 + \log\chi_{s}^{2} \right)
-k_{2}\left( \frac{\chi_{s}^{2}}{\chi_{s}^{2} + \tilde{v}^{2}} + \log( \chi_{s}^{2} + \tilde{v}^{2} ) \right) \right]\chi_{s} = 0.
\label{eq:stat2}
\end{equation}
Since $\tilde{\omega}=\omega/m$ the condition for the existence of PQ-ball is given by $0<\tilde{\omega}<1+2(k_{2}-k_{1})$. 

In conventional Q-ball models, the boundary condition was given by equation (\ref{eq:bc1}). However, in the present PQ-ball model, the field asymptotically approaches a finite positive value at infinity $\rho\rightarrow\infty$. Therefore, we modify the boundary conditions as follows:
\begin{equation}
	\left.\frac{d\phi}{d\rho}\right|_{\rho\rightarrow0} = 0, \ \ 
	\left.\frac{d\phi}{d\rho}\right|_{\rho\rightarrow\infty} = 0.
	\label{eq:bc2}
\end{equation}
The field magnitude at the origin, $\chi_{s0}\equiv\chi_{s}(\rho=0)$, and its asymptotic value at infinity, $\chi_{s\infty}\equiv\chi_{s}(\rho\rightarrow\infty)$, must be determined numerically. Since the initial and final velocities are given by the boundary conditions (\ref{eq:bc2}), specifying $\chi_{s0}$ uniquely determines $\chi_{s\infty}$. 
To find a solution under these conditions, the shooting method is adopted in our simulations.

\begin{figure}[t]
\centering
\includegraphics[width=1.0\linewidth]{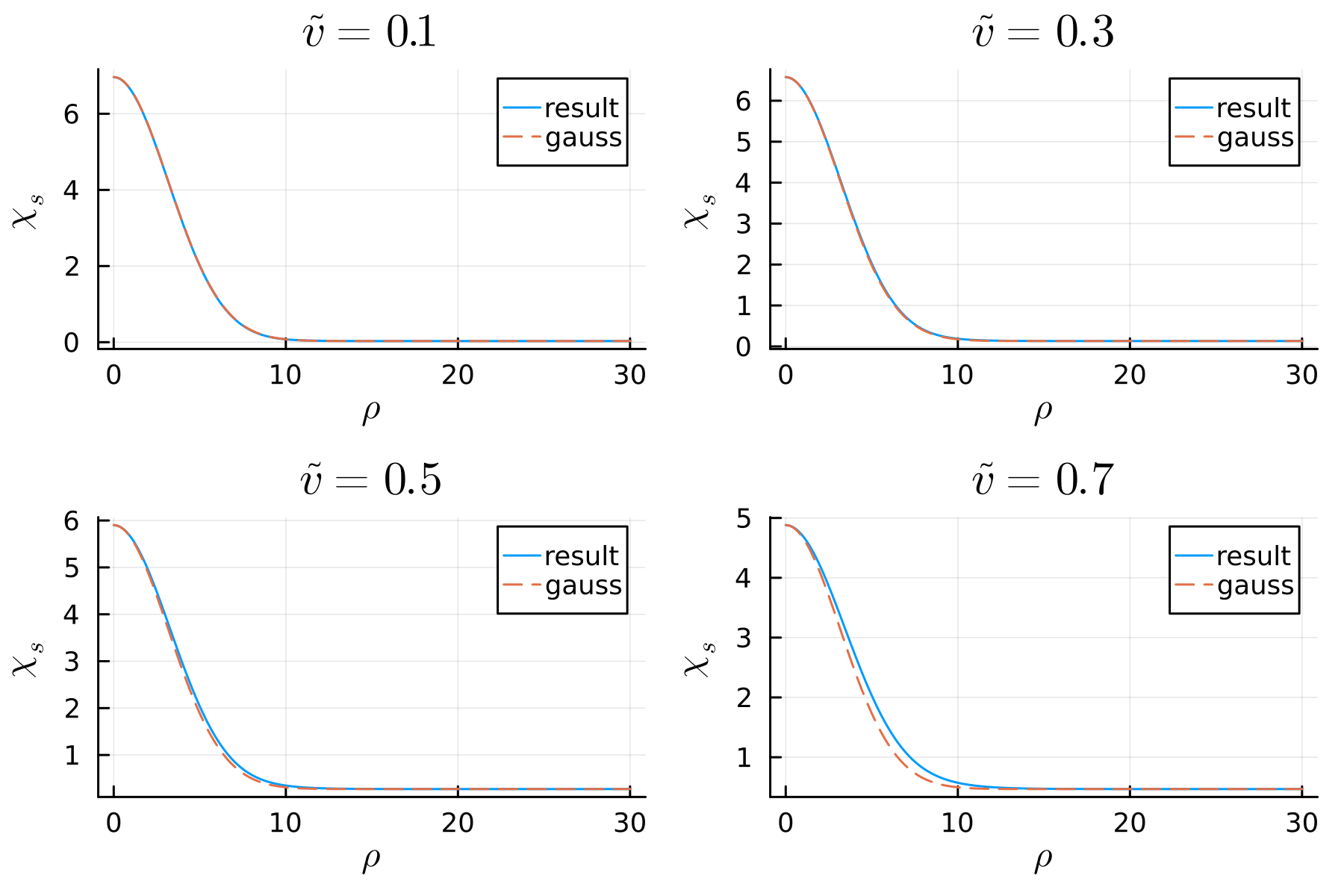}
\caption{A comparison of the numerically computed stationary solution (solid line) and the Gaussian function (dashed line) for $k_{1}=0.4,k_{2}=0.5,\tilde{\omega}=0.9$.}
\label{fig:gauss2}
\end{figure}

The stationary solution was computed by discretizing the spatial variable $\rho$ and replacing derivatives with finite differences. The grid spacing for $\rho$ was set to 0.01. The value of the field at $\rho=0$, $\chi_{s0}$ was determined using the shooting method applied to Eq.~(\ref{eq:stat2}). The results for
$k_{1}=0.4,k_{2}=0.5,\tilde{\omega}=0.9$ are shown in Fig.~\ref{fig:gauss2}. The figure illustrates the field profiles for fixed
$k_{1},k_{2},\tilde{\omega}$ while varying $\tilde{v}=0.1,0.3,0.5,0.7$. The solid blue line represents the computed radial field distribution. The shape is generally consistent with the theoretical calculation in the gravity-mediated model, except that in this case, the field asymptotically approaches a positive finite value instead of zero at infinity. 

The dashed orange line represents the Gaussian function:
\begin{equation}
\chi_{s}(r) = \chi_{s0}\exp\left(-\frac{r^{2}}{R_{Q}^{2}}\right) + \chi_{s\infty},
\end{equation}
where $R_\text{Q}$ is the Gaussian radius, defined by
\begin{equation}
	R_{Q} \simeq \frac{\sqrt{2}}{m\sqrt{k_{2}-k_{1}}}.
    \label{eq:gauss_rad}
\end{equation}
From Fig.~\ref{fig:gauss2}, we see that for small $\tilde{v}$, the computed results match well with the Gaussian function, but as $\tilde{v}$ increases, deviations become apparent, as expected.

\subsection{Time evolution of charges}
\label{sec:evolution}

\begin{figure}[t]
\centering
\includegraphics[width=0.7\linewidth]{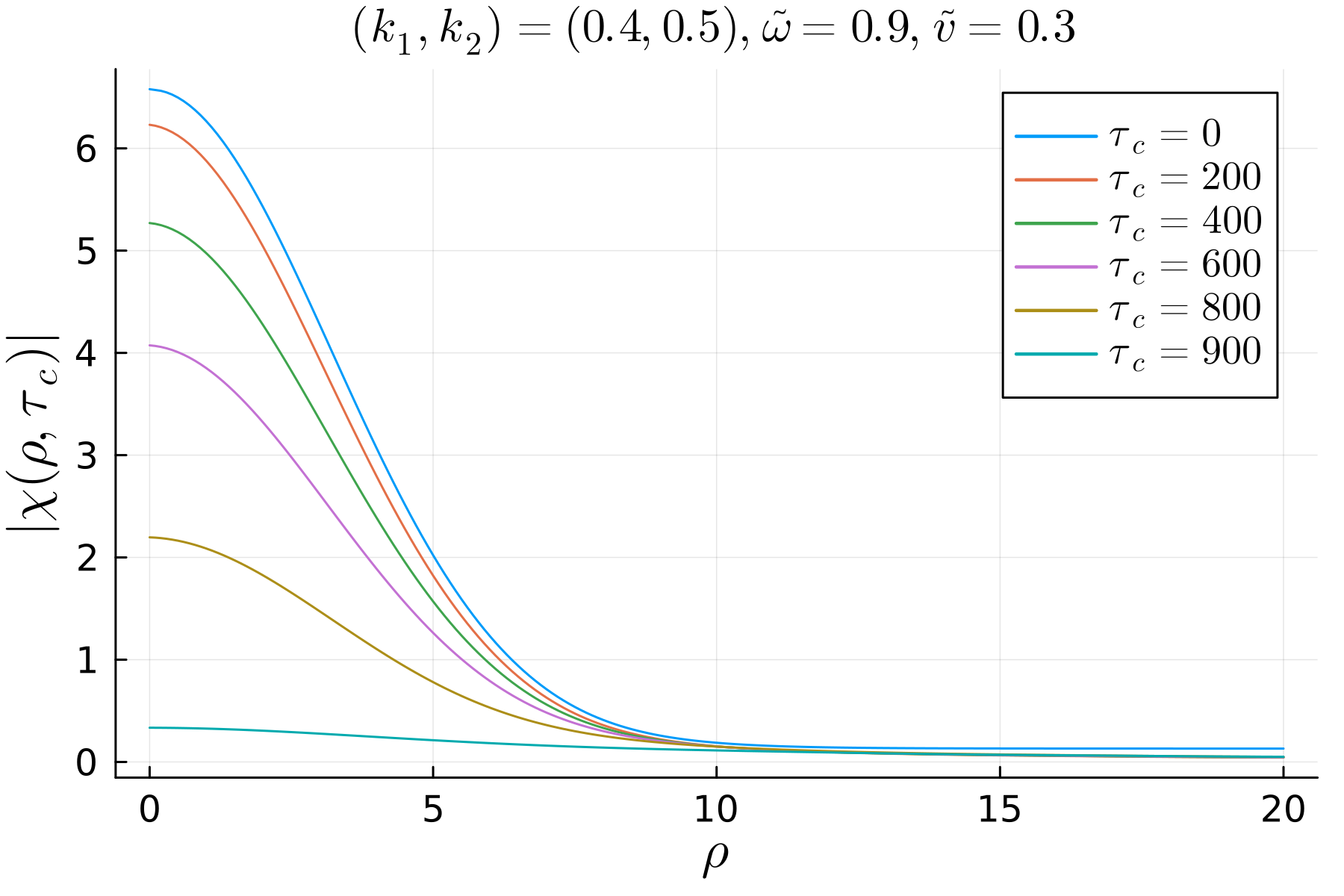}
\caption{Time evolution of $|\chi(\rho,\tau)|$ for $k_{1}=0.4,k_{2}=0.5,\tilde{v}=0.3,\tilde{\omega}=0.9$. 
We take $\tau = \tau_c = 0, 200, 400, 600, 800, 900$. 
The field magnitude decreases monotonically in the central region. The absorbing boundary condition is applied at $\rho=20$.}
\label{fig:changefield1}
\end{figure}

Now let us study the time evolution of the PQ-ball solution.
The time evolution of $\Phi$ is governed by Eq.~(\ref{eq:EL3}). The initial conditions at $\tau=0$ are given by
\begin{equation}
\chi|_{\tau=0} = \chi_{s}(\rho), \ \ \ \ \dot{\chi}|_{\tau=0} = i\tilde{\omega}\chi_{s}(\rho)
\end{equation}
where $\dot{\chi}$ represents the derivative with respect to $\tau$, and $\chi_{s}(\rho)$ is the stationary solution obtained in Sec.~\ref{static_solution}. Expressing the complex scalar field $\chi$ in terms of two real scalar fields,
$\chi(\rho,\tau)=\chi_{1}(\rho,\tau)+i\chi_{2}(\rho,\tau)$, the initial conditions can be rewritten as
\begin{equation}
\chi_{1}|_{\tau=0} = \chi_{s}(\rho), \ \ \ \ \dot{\chi}_{1}|_{\tau=0} = 0
\label{eq:inicond1}
\end{equation}
\begin{equation}
\chi_{2}|_{\tau=0} = 0, \ \ \ \ \dot{\chi}_{2}|_{\tau=0} = \tilde{\omega}\chi_{s}(\rho)
\label{eq:inicond2}
\end{equation}
The stationary solution obtained in Sec.~\ref{static_solution} describes a field rotating with a uniform angular velocity $\omega$. The initial conditions above reflect this rotational motion.

We apply the absorbing boundary conditions described in App.~\ref{sec:boundary}, in order to mimic the situation in which the environment around the PQ-ball is the vacuum.
We numerically calculated the time evolution of the field. Using a time step $\Delta\tau=0.02$ and a spatial step $\Delta\rho=0.1$, Eq.~(\ref{eq:EL3}) was solved using the fourth-order Leap-Frog method. 
Fig.~\ref{fig:changefield1} shows the spatial distribution of the field magnitude $|\chi|=\sqrt{\chi_{1}^{2}+\chi_{2}^{2}}$ 
at various times from $\tau=0$ to $\tau=900$ for 
$k_{1}=0.4,k_{2}=0.5,\tilde{\omega}=0.9,\tilde{v}=0.3$.
The field magnitude decreases monotonically in the central region over time.

\begin{figure}[t]
\centering
\includegraphics[width=1.0\linewidth]{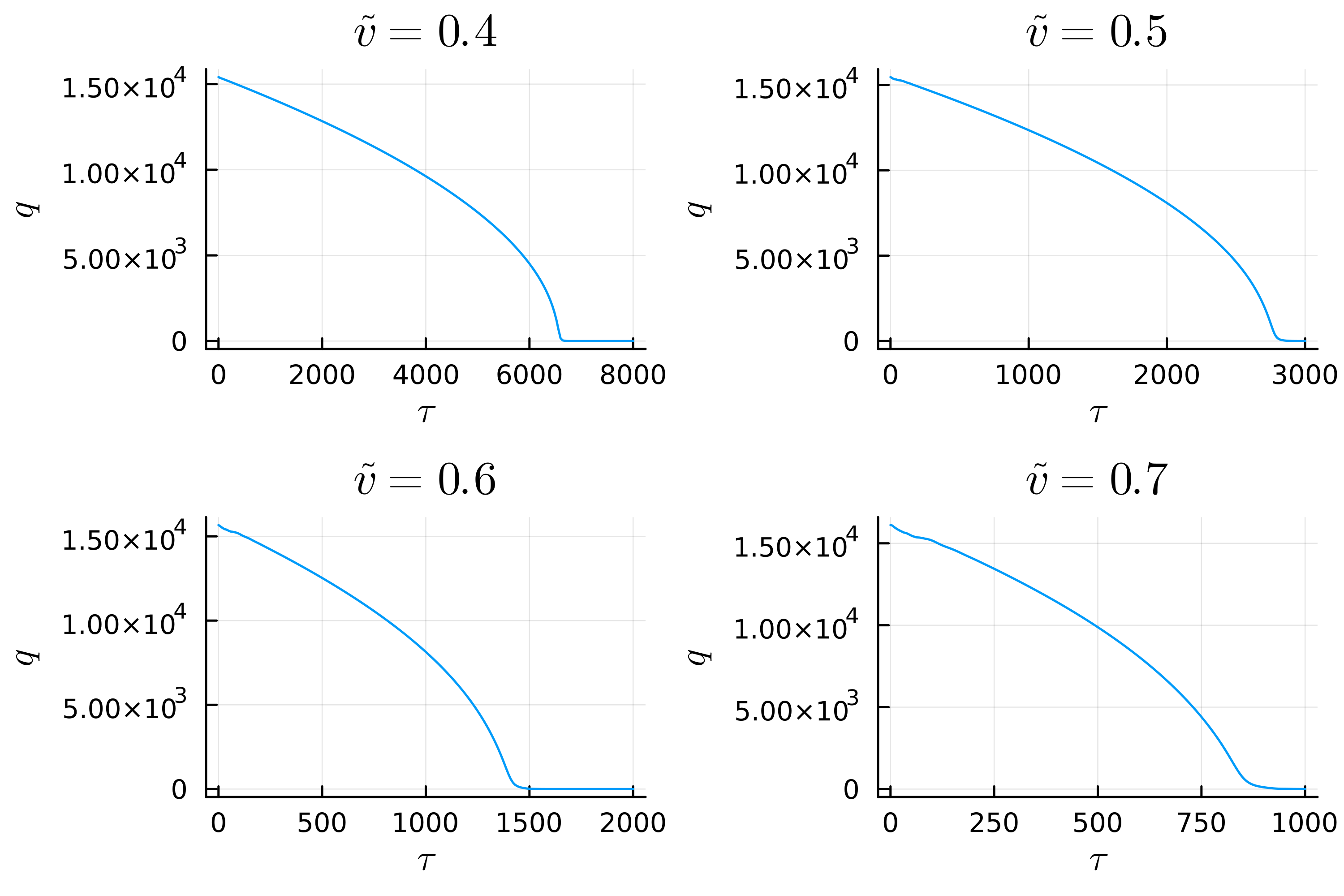}
\caption{Time evolution of the charge $q$ and its dependence on $\tilde{v}$ for $k_{1}=0.1,k_{2}=0.2,\tilde{\omega}=0.9$. 
}
\label{fig:Qchange1}
\end{figure}

Based on the stationary solution $\chi_{s}(\rho)$ we define the PQ-ball radius $R$ as the radial coordinate satisfying
\begin{equation}
\Phi_{s}(R)-\Phi_{s\infty} = 0.1\times(\Phi_{s0}-\Phi_{s\infty}).
\end{equation}
In numerical calculations, the integration range for charge $q$ is set between three to four times $\tilde{R}=mR$. 

Fig.~\ref{fig:Qchange1} shows the time evolution of $q$ for $k_{1}=0.1,k_{2}=0.2$ with different values of $\tilde{v}$. In all cases, $q$ decreases monotonically and eventually converges to zero. At early times, 
$q$ decreases linearly, but as 
$q$ becomes sufficiently small, the decay rate accelerates.

Focusing on the initial linear decay region in Fig.~\ref{fig:Qchange1}, we define the decay rate as the change in $q$ over a given time interval:
\begin{equation}
\frac{dq}{dt} = \frac{q(t_{2})-q(t_{1})}{t_{2}-t_{1}}.
\end{equation}
We investigate how the decay rate changes as $\tilde{v}$ varies while keeping $k_{1}$, $k_{2}$, and $\tilde{\omega}$ fixed. 

\begin{figure}[t]
\centering
\includegraphics[width=1.0\linewidth]{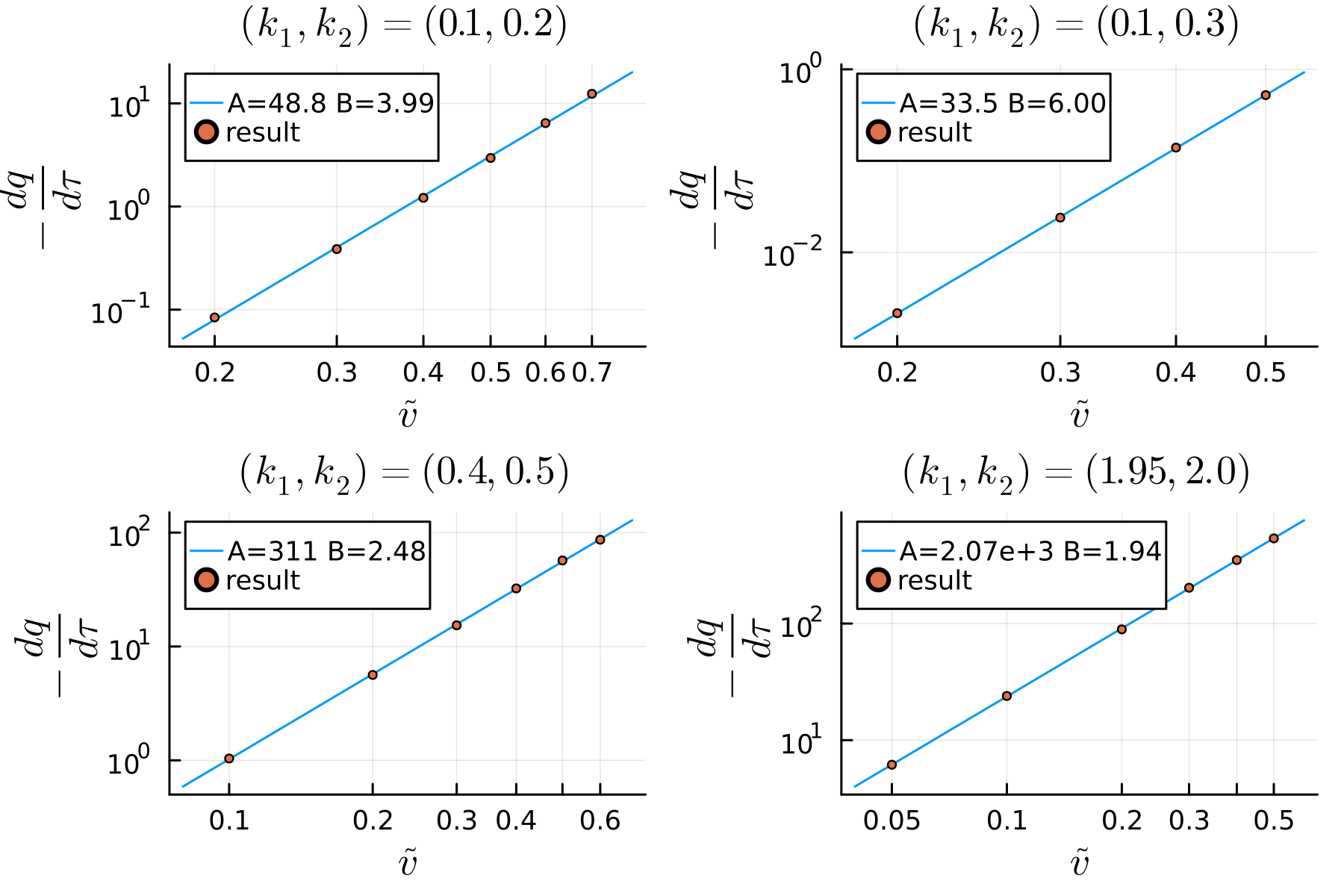}
\caption{Dependence of the decay rate $dq/dt$ on $\tilde{v}$ for 
$\tilde{\omega}=0.9$ with different values of $(k_{1},k_{2})$. The points represent numerical results, and the lines correspond to the fitted function $y=Ax^{B}$.}
\label{fig:dQdtchange1}
\end{figure}

Fig.~\ref{fig:dQdtchange1} shows the numerical results for the dependence of $dq/dt$ for $(k_{1},k_{2})= (0.1,0.2)$, $(0.1,0.3)$, $(0.4,0.5)$, $(1.95,2.0)$. 
In all cases, the data points align approximately along a straight line in a log-log plot, suggesting that the decay rate can be approximated as a power law: 
\begin{equation}
\frac{dq}{dt} = -Av^{B}.
\end{equation}
By fitting the data, we find that the exponent $B$ follows the empirical relation:
\begin{equation}
B = 2\frac{k_{2}}{k_{1}}.
\label{eq:paradep}
\end{equation}

This parameter dependence can be simplified as follows. As discussed in Sec.~\ref{sec:PQQball}, the field value at infinity, $\Phi_{s\infty}$ corresponds to the local maximum of the potential $U_\omega(|\Phi|)$ in Fig.~\ref{fig:Uax}. Thus, the smallest nonzero extremum of the potential represents the asymptotic value of $\Phi_{s}$ at infinity. The stationary solution satisfies:
\begin{equation}
-\tilde{\omega}^{2} + 1 + k_{1}\left( 1 + \log( \chi_{s}^{2} ) \right)-k_{2}\left( \frac{\chi_{s}^{2}}{\chi_{s}^{2} + \tilde{v}^{2}} + \log( \chi_{s}^{2} + \tilde{v}^{2} ) \right) = 0.
\end{equation}
Assuming $\chi_{s\infty}\ll\tilde{v}$, we approximate:
\begin{equation}
k_{1}\log(\chi_{s\infty}^{2}) - k_{2}\log(\tilde{v}^{2}) = \tilde{\omega}^{2} - 1 -k_{1}.
\label{eq:chi-v}
\end{equation}
Eq.~(\ref{eq:chi-v}) is rewritten as
\begin{equation}
	\chi_{s\infty} = \tilde{v}^{\frac{k_{2}}{k_{1}}}\exp\left(\frac{1}{2k_{1}}(\tilde{\omega}^{2}-1-k_{1})\right).
	\label{eq:chi_inf}
\end{equation}
Therefore, the result of Eq.~(\ref{eq:paradep}) indicates
\begin{equation}
\frac{dq}{dt} \sim v^{\frac{2k_{2}}{k_{1}}} \propto \Phi_{s\infty}^{2}.
\end{equation}
This is consistent with theoretical interpretation, as we will see in the next subsection.

\begin{figure}[t]
\centering
\includegraphics[width=1.0\linewidth]{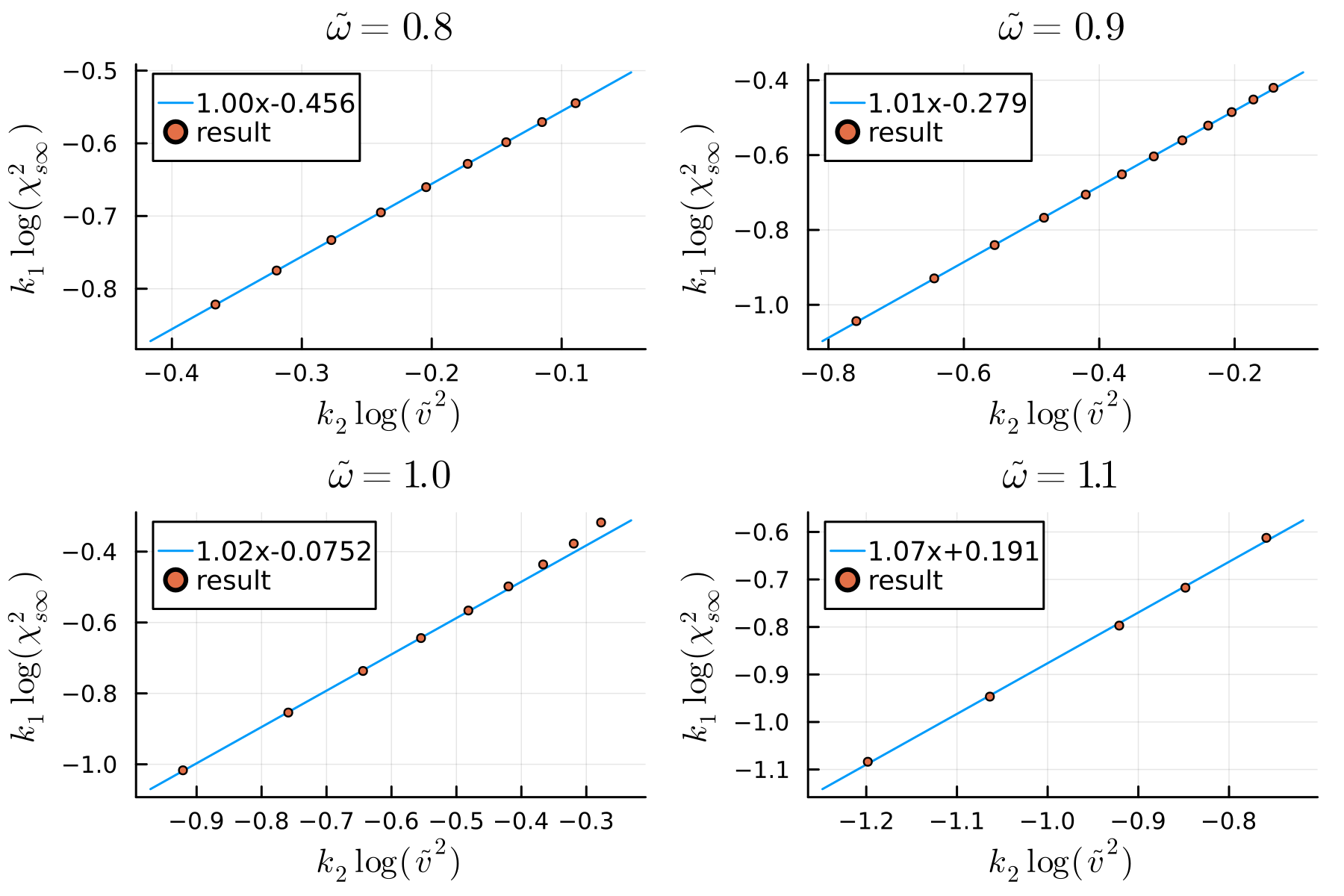}
\caption{The relationship between $\chi_{s\infty}$ and $\tilde{v}$ 
when varying $\tilde{\omega}$. The case of 
$(k_{1},k_{2})=(0.1,0.2)$  is shown. The points represent numerical results, while the straight line corresponds to a linear fit $y=Cx$ As $\tilde{v}$ increases, $\chi_{s\infty}/\tilde{v}$ becomes large, and the linear approximation breaks down. The nonlinear behavior at the upper right for $\tilde{\omega}=1.0$ illustrates this, and in this particular graph, the four rightmost points were excluded from the fitting.}
\label{fig:vchi}
\end{figure}

Before doing so, we verify the relationship between $\tilde{v}$ and $\chi_{s\infty}$ using numerical calculations. Fig.~\ref{fig:vchi} plots the numerical results, with the horizontal axis representing $k_{2}\log(\tilde{v}^{2})$ and the vertical axis representing $k_{1}\log(\chi_{s\infty}^{2})$. A clear positive correlation is observed between the two values, and the slope is approximately 1, which is consistent with Eq.~(\ref{eq:chi-v}). For small $\tilde{\omega}$, the intercept position ($y=\tilde{\omega}-1-k_{1}$) also agrees well with Eq.~(\ref{eq:chi-v}). 
However, when $\tilde{\omega}$ and $\tilde{v}$ become large, the linear relationship begins to break down. This is likely due to $\chi_{s\infty}/\tilde{v}$ approaching 1, which reduces the accuracy of the assumed approximation.

\subsection{Decay rate formula}

From the results of the previous subsection, we found that the decay rate  $dQ/dt$ is proportional to $\Phi_{s\infty}^2$. It may suggest that decay rate is proportioanl to the charge density near the PQ-ball surface, given by $\omega\Phi_{s\infty}^{2}$. Since the PQ-ball decay rate corresponds to the total flux of charge passing through the surface per unit time, it should also be proportional to the surface area $4\pi R^{2}$. Therefore, it can be expressed, up to an undetermined constant factor, as
\begin{equation}
	\frac{dQ}{dt} \sim -4\pi v_{q}\omega R^{2}\Phi_{s\infty}^{2},
	\label{eq:decayrate}
\end{equation}
where $v_{q}$ represents the velocity of the outgoing charge flow.\footnote{
A similar result had also been obtained in a Q-ball model that includes interactions with other particles~\cite{Cohen:1986ct}:
they demonstrated that the decay rate is proportional to the Q-ball surface area.}
By using the Gaussian radius $R_{Q}$ from Eq.~(\ref{eq:gauss_rad}) as the PQ-ball radius $R$, and substituting $\omega^2 \sim \left(1+2(k_2-k_1)\right)m^2$, the decay rate is given by: 
\begin{equation}
	\frac{dQ}{dt} \sim -4\pi\sqrt{1+2(k_{2}-k_{1})}m\times\frac{2}{m^{2}(k_{1}-k_{2})}\times M^{2}\left( \frac{v}{m} \right)^{\frac{2k_{2}}{k_{1}}}\exp\left(\frac{1}{k_{1}}(2k_{2}-3k_{1})\right),
	\label{eq:decayrate2}
\end{equation}
where we assume the charge flow velocity is relativistic $v_{q}=1$. This expresses Eq.~(\ref{eq:decayrate}) in terms of the fundamental parameters $(k_{1},k_{2},v,m,M)$ of the original potential (\ref{eq:potential_PQ}). Eqs.~(\ref{eq:decayrate}) and (\ref{eq:decayrate2}) represent the primary results of this study.

\begin{figure}
\centering
\includegraphics[width=0.6\linewidth]{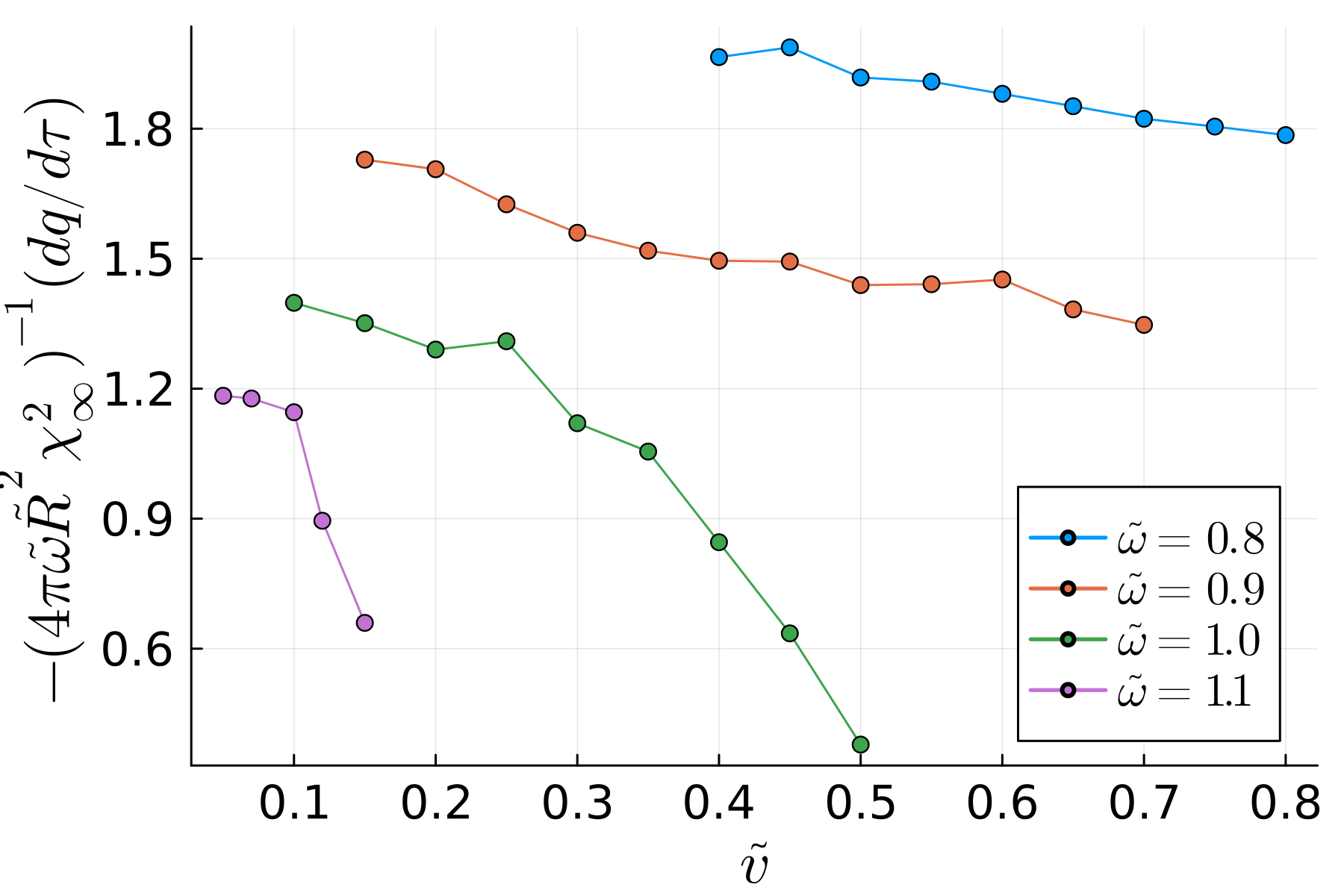}
\includegraphics[width=0.6\linewidth]{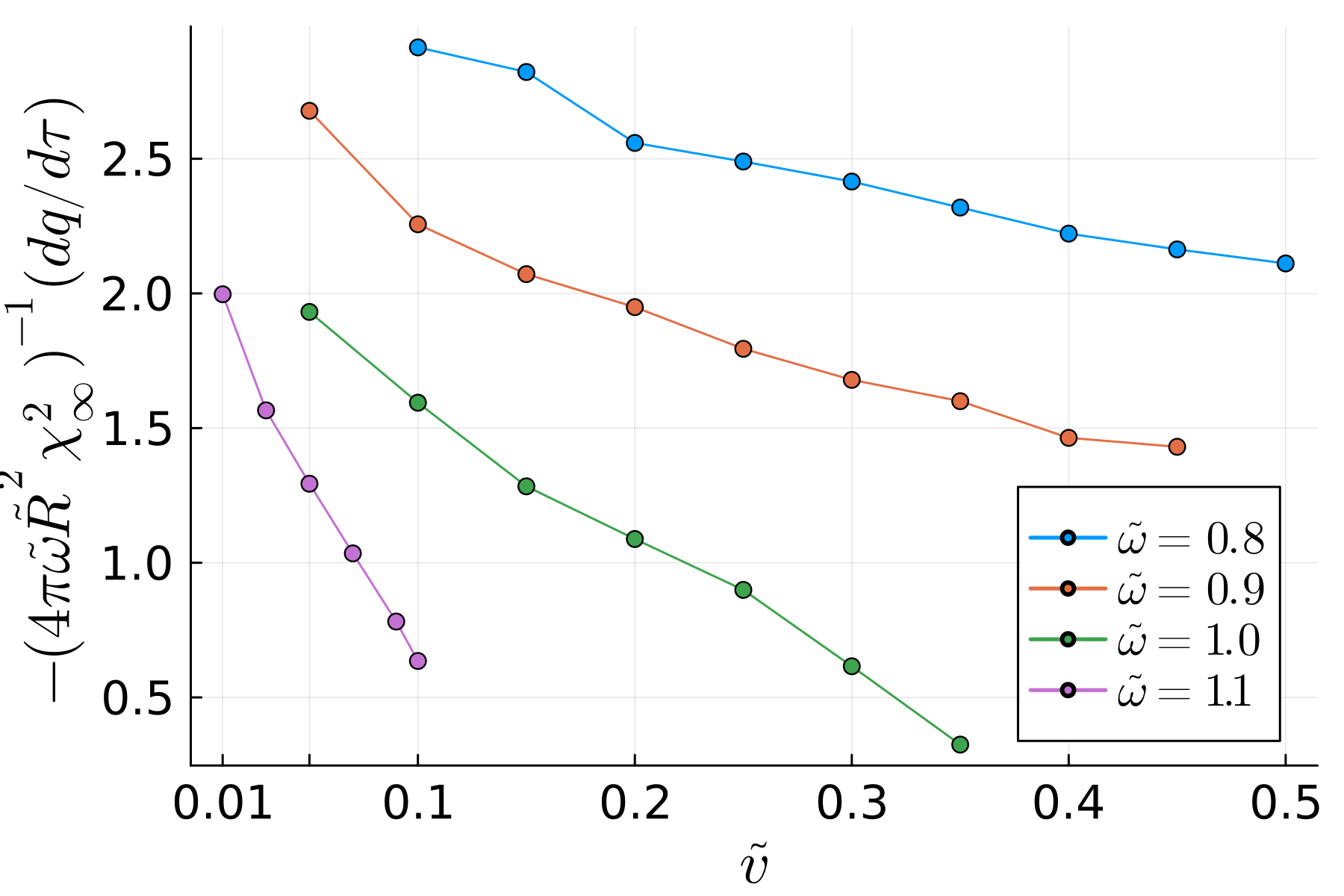}
\caption{The ratio between the numerically computed PQ-ball decay rate and the charge flux from the PQ-ball surface. The ratio was computed for different values of $\omega$ and $v$, assuming $v_{q}=1$. 
The upper graph shows the case of $k_{1}=0.1,k_{2}=0.2$, while the lower graph corresponds to 
$k_{1}=0.4,k_{2}=0.5$.}
\label{fig:decayrate}
\end{figure}

We numerically verify the validity of (\ref{eq:decayrate}). 
Fig.~\ref{fig:decayrate} shows the ratio of the left-hand side to the right-hand side of equation (\ref{eq:decayrate}) for $(k_{1},k_{2})=(0.1,0.2),(0.4,0.5)$ computed for different values of $\omega$ and $v$. Although the ratio decreases with increasing $\omega$ and $v$, it remains approximately within $\mathcal{O}(1)$. Therefore, the theoretical prediction given by Eq.~(\ref{eq:decayrate}) is valid within an uncertainty of $\mathcal{O}(1)$.

Let us reconsider the physical meaning of our findings. The stationary solution of a PQ-ball takes a finite asymptotic value 
$\Phi=\Phi_{s\infty}e^{i\omega t}$ at infinity, indicating that it corresponds to a stable solution in a finite-density environment. In such a scenario, there should be an inflow of charge from the surrounding environment into the PQ-ball. However, since the solution is stationary, there must also be an outgoing charge flow from the PQ-ball to the surrounding environment, and the two fluxes must balance so that no net charge change occurs.
When the finite-density environment is removed, the inflow of charge into the PQ-ball ceases, leaving only the outgoing flux. Consequently, the charge of the PQ-ball is expected to decrease irreversibly. Based on this interpretation, the decay rate of the PQ-ball should be proportional to both the surface area of the PQ-ball and the charge density near the surface, $\sim\omega\Phi_{s\infty}^{2}$. This provides a physical explanation for the decay behavior observed in our numerical simulations.

It is particularly important to emphasize that the decay rate depends on the asymptotic field value at infinity, 
$\Phi_{s\infty}$, rather than the vacuum expectation value $f_{a}$. Depending on models, 
$f_{a}$ and $\Phi_{s\infty}$ can differ by several orders of magnitude, making the distinction between the two scales physically and numerically significant.

\section{Applications}
\label{sec:app}

In this section we discuss some cosmological implications of the PQ-balls with decay rate obtained in the previous section.

Using Eq.~(\ref{eq:decayrate2}), we can estimate the decay time of the PQ-ball. The total charge of the PQ-ball can be approximated as:
\begin{equation}
	Q = \int\omega\Phi_{s}^{2}d^{3}x= 4\pi\omega\Phi_{s0}^{2}\int^{\infty}_{0}r^{2}e^{-\frac{2r^{2}}{R^{2}}}dr \simeq \left(\frac{\pi}{2}\right)^{\frac{3}{2}}\omega R^{3}\Phi_{s0}^{2}.
\end{equation}
The approximate time required for the PQ-ball to completely decay can be estimated as the inverse of the quantity obtained by dividing the decay rate by the total charge:
\begin{equation}
	\left|\frac{1}{Q}\frac{dQ}{dt}\right| \sim \frac{8\sqrt{2}}{\sqrt{\pi}R}\left(\frac{\Phi_{s\infty}}{\Phi_{s0}}\right)^{2} = \frac{8}{\sqrt{\pi}}m\sqrt{k_{2}-k_{1}}\left(\frac{\Phi_{s\infty}}{\Phi_{s0}}\right)^{2}.
	\label{eq:rateratio}
\end{equation}
Note that the ratio $\Phi_{s\infty}/\Phi_{s0}$ can be much smaller than unity and PQ-balls can be very long-lived.
By comparing this quantity with the Hubble parameter, we can estimate the epoch at which the PQ-ball decays and disappears.
Assuming that the decay happens at the radiation-dominaed Universe, the decay temperature $T_{\rm dec}$ is estimated as
\begin{equation}
	T_{\rm dec} \simeq \mathcal C \times (k_2-k_1)^{\frac{1}{4}} \sqrt{m M_{\rm Pl}}\left(\frac{\Phi_{s\infty}}{\Phi_{s0}}\right),
    \label{eq:T_dec}
\end{equation}
where $M_{\rm Pl}$ denotes the reduced Planck scale and $\mathcal C = \left(\pi^2 g_* /5760\right)^{-1/4}$ is $\mathcal O(1)$ constant with $g_*$ being the relativistic degrees of freedom at $T_{\rm dec}$.

As an example, let us work with a specific parameter choice:
$(k_{1},k_{2})=(0.1,0.2)$, from which we obtain $\tilde{\omega}=\sqrt{1.2}\simeq1$. 
Additionally, we set the following values for the other constant parameters:
\begin{equation}
	M = 10^{16}\,\text{GeV},\ \ m = 1\,\text{TeV},\ \ v = 10^{15}\,\text{GeV}.
\end{equation}
Thus, $\tilde{v}=10^{-1}$, which is small enough that the approximations used in the previous calculations remain valid.
The energy scale of spontaneous symmetry breaking, $\sim f_{a}$, corresponds to the position of the potential minimum in Fig.~\ref{fig:Vax1}. While its exact value must be determined numerically, when 
$f_{a}\ll v$, it can be approximated as
\begin{equation}
	\frac{f_{a}}{M} = \exp\left(\frac{-1-k_{1}}{2k_{1}}\right)\left( \frac{v}{M} \right)^{\frac{k_{2}}{k_{1}}},
\end{equation}
which allows us to determine 
$f_{a}$ in terms of 
$k_{1}$, $k_{2}$, $M$, and $v$. Substituting the above parameter values, we find:
\begin{equation}
	f_{a} \sim e^{-11/2}\times10^{-2}\times10^{16}\text{GeV} \simeq 4\times10^{11}\,\text{GeV}.
\end{equation}
From Eq.~(\ref{eq:chi_inf}), the asymptotic field value at infinity is given by
\begin{equation}
	\Phi_{s\infty} = M\times\tilde{v}^{2}\times e^{\frac{1}{2}} \simeq 10^{-2}M.
\end{equation}
Assuming that $\Phi_{s0}$ is of the same energy scale as $M$, we obtain $\Phi_{s\infty}/\Phi_{s0}\sim10^{-2}$. 
Substituting these values into Eq.~(\ref{eq:rateratio}), we obtain
\begin{equation}
	\frac{1}{Q}\frac{dQ}{dt} \sim \left(\frac{\Phi_{s\infty}}{\Phi_{s0}}\right)^{2}m
	= 100\,\text{MeV}.
\end{equation}
From Eq.~(\ref{eq:T_dec}), the decay temperature is estimated as
\begin{equation}
	T_\text{dec} \simeq 3\times10^{8}\,\text{GeV}.
\label{eq:dectemp}
\end{equation}
From the numerical results, as discussed in the previous section, whether the decay rate is proportional to $f_{a}^{2}$ or $\Phi_{s\infty}^{2}$
leads to a significant difference of about $(f_{a}^{2}/\Phi_{s\infty}^{2})\sim10^{-5}$. 
This results in a temperature difference of about $10^{2.5}$ times at the decay epoch, causing major differences in cosmological predictions. Thus, the formula (\ref{eq:decayrate}) obtained here provides a numerically nontrivial result.

Revisiting the cosmological scenario in the context of the kinetic misalignment mechanism, the following picture may emerge. A mechanism similar to the Affleck-Dine mechanism leads to the rotation of the PQ-charged field 
$\Phi$ in the phase direction from a large vacuum expectation value, generating a net 
$\text{U}(1)_\text{PQ}$ charge. Depending on the shape of the potential for 
$\Phi$, density fluctuations in the charge distribution grow, leading to the formation of PQ-balls. During this formation process, most of the
$\text{U}(1)_\text{PQ}$ charge is absorbed into the PQ-balls. Later, at the time corresponding to the temperature given by Eq.~(\ref{eq:T_dec}), the PQ-balls completely decay.
The final abundance of axions is determined by the total amount of generated
$\text{U}(1)_\text{PQ}$ charge and is expected to be unaffected by the formation and decay of PQ-balls, as explained in Appendix~\ref{sec:kinetic}. However, the dissipation of the radial energy of $\Phi$ is expected to be significantly altered by PQ-ball formation. 
Since interactions with other particles occur primarily near the PQ-ball surface, most of the particles absorbed into the PQ-ball are likely to survive without changing species until they are emitted.

In general, the saxion is more massive than the axion and has decay channels into two axions~\cite{Chun:1995hc}. 
If saxions remain in the universe for a relatively long time, they will eventually decay and release high-energy axions. These axions contribute to the radiation component, and an excessive amount could exceed the observed radiation density, posing a potential problem~\cite{Choi:1996vz,Hashimoto:1998ua,Asaka:1998xa,Ichikawa:2007jv,Kawasaki:2007mk}.
However, saxions can interact with the thermal bath in the early universe and transfer their energy to the Standard Model particles and other species, the rate of which is characterized by thermal dissipation rate~\cite{Yokoyama:2004pf,Bastero-Gil:2010dgy,Drewes:2010pf,Mukaida:2012qn,Mukaida:2012bz,Drewes:2013iaa,Mukaida:2013xxa,Moroi:2014mqa,Tanin:2017bzm}. This significantly reduces both the number and energy of saxions, potentially resolving the issue. This process is referred to as saxion thermalization.
Detailed analysis of saxion thermalization in the context of kinetic misalignment is found in Refs.~\cite{Co:2020dya,Eroncel:2024rpe}.
The conditions required for thermalization depend significantly on the state of saxions during this process. 
In particular, there is a crucial difference between scenarios where saxions are homogeneously distributed throughout space and scenarios where they first form PQ-balls before decaying. These differences affect both the timing and degree of thermalization due to the presence of PQ-balls, leading to vastly different cosmological implications.
This could lead to a revision of the allowed parameter space in the kinetic misalignment mechanism, which was previously constrained by the limits on dark radiation originating from saxions.

\section{Discussions and conclusion}
\label{sec:discussions}

We have demonstrated that Q-balls carrying a 
U(1)$_{\rm PQ}$ charge can form in certain types of PQ models, which we call PQ-balls, and have numerically investigated their formation and decay behavior. Theoretically, PQ-balls exist as stable solutions in a finite-density environment. Therefore, such PQ-balls are expected to be unstable in a vacuum. However, the precise manner of their decay is not immediately obvious. In this study, we imposed absorbing boundary conditions at a sufficiently large radius under the assumption of spherical symmetry and numerically evolved the discretized equations of motion over time. The results showed that the PQ-balls gradually decayed. 
Furthermore, it was confirmed that the charge depletion rate is proportional to the square of the field magnitude at infinity for the stable solution in a hypothetical finite-density medium. This suggests that the decay rate is approximately proportional to the charge density at the PQ-ball surface. Since the rate of charge reduction corresponds to the total flux passing through the PQ-ball surface per unit time, our numerical results suggest that it can be effectively expressed by multiplying the charge density at the surface by the velocity of outgoing particles, as given in Eq.~(\ref{eq:decayrate}). This was numerically verified to hold with an uncertainty of $O(1)$. Since Eq.~(\ref{eq:decayrate}) provides a physically natural interpretation, it is expected to be applicable not only to the specific model analyzed in this study but also to more general cases where Q-balls are formed by a complex scalar field with a spontaneously symmetry-breaking potential.
What is phenomenologically important is that the PQ-ball lifetime can be extremely long if the ratio $\Phi_{s\infty}/\Phi_{s0}$ is small enough (see Eq.~(\ref{eq:rateratio})).
Actually it can take any small value, and it ensures the importance of PQ-balls as they are meta-stable objects in cosmological time scales.

Our results can be applied to promising axion production mechanisms. For example, in the kinetic misalignment mechanism, a dynamical process similar to the Affleck-Dine mechanism occurs, leading to the potential formation of PQ-balls. However, whereas in the Affleck-Dine mechanism Q-balls carry a U(1) charge corresponding to baryon or lepton number, in the kinetic misalignment mechanism, the resulting Q-balls carry a PQ charge. 
The radial degrees of freedom of the complex scalar field forming these PQ-balls possess large energy, and the timing of energy transfer to axions or the Standard Model particles significantly affects the predicted cosmological scenario.
Thus, determining when such PQ-balls form and how long they take to decay is of great importance. The decay rate formula derived in this study provides a useful tool for analyzing these questions in detail.

Finally, we note that it is nontrivial to understand how PQ-balls emerge in an expanding Universe, and what their structure looks like throughout the thermal history. Addressing these issues may require dedicated lattice simulations, although we have confirmed their formation in a spherically symmetric system within static spacetime. We leave a detailed investigation for future work.

\ 

{\it {\bf Note added:} 
At an early stage of this work, we became aware that another group~\cite{accompanypaper} was independently investigating a related topic, specifically, the stability of Q-ball-like objects in a model with spontaneously broken U(1) symmetry.
We agreed to pursue our research independently while coordinating to submit our papers on the same day.
Whereas we study a single-field model, their approach involves a two-field setup: one field is responsible for Q-ball formation, while the other field is associated with spontaneous symmetry breaking.
}

\section*{Acknowledgements}
We thank Kai Murai and Fuminobu Takahashi for useful discussion.
This work was supported by JSPS KAKENHI Grant Numbers 24K07010 [KN],
20H05851 [MY], 23K13092 [MY].
This work was supported by World Premier International Research Center Initiative (WPI), MEXT, Japan.

\appendix

\section{Linear perturbartion theory for PQ-ball formation}
\label{sec:linear}

In this Appendix, we discuss the formation mechanism of (P)Q-balls. We consider a scenario in which a spatially homogeneous condensate with a conserved U(1) charge exists due to some mechanism in the early universe. For example, in the kinetic misalignment mechanism, the U(1) charge corresponds to the PQ charge. Although such a field configuration is almost uniform in space, the impact of small density fluctuations in the universe cannot be ignored. In some cases, the instability of these fluctuations can lead to regions of higher and lower charge densities. As these inhomogeneities grow, large clumps of charge may form, eventually leading to the formation of (P)Q-balls~\cite{Enqvist:1998en, Kusenko:1997si, Enqvist:1997si}. Here, we investigate the growth of these fluctuations in the linear approximation, assuming they are initially small.

First, we decompose the complex scalar field into its radial and angular components such as 
\begin{equation}
\Phi = P e^{i\Omega}
\end{equation}
where 
$P$ and $\Omega$ are real scalar fields that depend on both time and space. The potential $U$ has U(1) symmetry, meaning it can be written as a function of 
$P$ alone, $U(\Phi)=U(P)$. Under the spherically symmetric metric, the classical equations of motion for $P$ and $\Omega$ take the form of 
\begin{equation}
\ddot{\Omega} + 3H\dot{\Omega} - \frac{1}{a^{2}(t)}\Delta\Omega + \frac{2\dot{P}}{P}\dot{\Omega}
- \frac{2}{a^{2}(t)P}(\partial_{i}\Omega)(\partial^{i}P) = 0
\end{equation}
\begin{equation}
\ddot{P} + 3H\dot{P} - \frac{1}{a^{2}(t)}\Delta P - \dot{\Omega}^{2}P
+ \frac{1}{a^{2}(t)}(\partial_{i}\Omega)^{2}P + \frac{\partial U}{\partial P} = 0
\end{equation}
Here, dots denote time derivatives, and the italic subscript $i$ runs from 1 to 3 representing spatial components. The scale factor is denoted by 
$a(t)$, and the Hubble parameter is defined by $H\equiv\dot{a}/a$. 
We now consider small perturbations around a spatially homogeneous background field such as 
\begin{align}
P(\vec{x},t) &= P(t) + \delta P(\vec{x},t) \\
\Omega(\vec{x},t) &= \Omega(t) + \delta\Omega(\vec{x},t)
\end{align}
Substituting these into the equations of motion and neglecting second-order terms in fluctuations, we obtain the equations for the perturbations $\delta P$ and $\delta\Omega$ such as 
\begin{equation}
\ddot{\delta\Omega} + 3H(\dot{\delta\Omega}) - \frac{1}{a^{2}(t)}\Delta(\delta\Omega)
+ \frac{2\dot{P}}{P}\dot{\delta\Omega} + \frac{2\dot{\Omega}}{P}(\dot{\delta P})
- \frac{2\dot{P}\dot{\Omega}}{P^{2}}\delta P = 0
\end{equation}
\begin{equation}
\ddot{\delta P} + 3H(\dot{\delta P}) - \frac{1}{a^{2}(t)}\Delta(\delta P)
- 2P\dot{\Omega}(\dot{\delta\Omega}) + U''\delta P -\dot{\Omega}^{2}\delta P = 0
\end{equation}

Since we consider small fluctuations, we assume their form as 
\begin{equation}
 \delta P,\delta\Omega\propto e^{S(t)-i\vec{k}\cdot\vec{x}}
\end{equation}
where we focus on growth modes with 
$\mathrm{Re}(\alpha)>0$ for $\alpha\equiv dS/dt$, describing their time evolution. Expressing time derivatives in terms of $\alpha$ and rewriting the equations in matrix form, we obtain 
\begin{equation}
\begin{pmatrix}
\dot{\alpha}+\alpha^{2}+3H\alpha+\frac{\vec{k}^{2}}{a^{2}}+\frac{2\dot{P}}{P}\alpha & -\frac{2\dot{P}\dot{\Omega}}{P^{2}}+\frac{2\alpha}{P}\dot{\Omega} \\
-2P\dot{\Omega}\alpha & \dot{\alpha}+\alpha^{2}+3H\alpha+\frac{\vec{k}^{2}}{a^{2}}-\dot{\Omega}^{2}+U''
\end{pmatrix}
\begin{pmatrix}
\delta\Omega \\
\delta P
\end{pmatrix}
=0
\end{equation}
For the perturbation to have a nontrivial solution, the determinant of this matrix must be zero, leading to the condition of 
\begin{equation}
\left[ \dot{\alpha}+\alpha^{2}+3H\alpha+\frac{\vec{k}^{2}}{a^{2}}+\frac{2\dot{P}}{P}\alpha \right]
\left[ \dot{\alpha}+\alpha^{2}+3H\alpha+\frac{\vec{k}^{2}}{a^{2}}-\dot{\Omega}^{2}+U'' \right]
+4\dot{\Omega}^{2}\alpha\left[ \alpha-\frac{\dot{P}}{P} \right] = 0
\end{equation}
Neglecting the terms proportional to $H$ (since they are small compared to 
$U''\sim m_{s}^{2}$), this simplifies to the quadratic equation in
\begin{equation}
(\alpha^{2})^{2} + \left( 2\frac{\vec{k}^{2}}{a^{2}}+3\dot{\Omega}^{2}+U'' \right)\alpha^{2}
+ \frac{\vec{k}^{2}}{a^{2}}\left( \frac{\vec{k}^{2}}{a^{2}}-\dot{\Omega}^{2}+U'' \right) = 0
\end{equation}
For instability, i.e., a growing mode $\alpha^{2}>0$, the necessary condition is
\begin{equation}
\frac{\vec{k}^{2}}{a^{2}}-\dot{\Omega}^{2}+U'' < 0
\end{equation}
which implies that the unstable modes must satisfy
\begin{equation}
0 < |\vec{k}| < a\sqrt{\dot{\Omega}^{2}-U''}
\end{equation}
The mode that grows fastest, i.e., the mode with maximum $\alpha$, corresponds to
\begin{equation}
\frac{\vec{k}_{\mathrm{max}}^{2}}{a^{2}} = \frac{1}{16\dot{\Omega}^{2}}\left( -U''^{2}-6U''\dot{\Omega}^{2}+7\dot{\Omega}^{4} \right)
\end{equation}
with the maximum growth rate given by
\begin{equation}
\alpha^{2} = \frac{1}{16}\left( -2U''+\frac{U''^{2}}{\dot{\Omega}^{2}}+\dot{\Omega}^{2} \right)
\end{equation}
Here, $U''$ appears explicitly, justifying our earlier assumption to not neglect 
$\alpha^{2}$. 
The amplitude of a perturbation mode with wave number $k$ at time 
$t$ (assuming it starts at $t=0$) is characterized by
\begin{equation}
S(k,t) = \int_{0}^{t}\alpha(k,t')dt'
\end{equation}
where the most unstable mode corresponds to $k=k_{\mathrm{max}}$
If $e^{S(k_{\mathrm{max}})}\gg 1$, the fluctuations grow significantly.

When the fluctuations become sufficiently large, specifically when 
$\langle\delta P^{2}\rangle\sim P^{2}$, or equivalently $S(k_{\mathrm{max}},t)\sim\log(P/\delta P)$, our linear perturbation analysis breaks down, and the charge distribution develops distinct high-density regions. As these high-density regions grow, they eventually evolve into (P)Q-balls.

Previous numerical studies~\cite{Kasuya:1999wu, Kasuya:2000wx} have confirmed that for Q-balls in the context of Affleck-Dine mechanism, the dominant mode just before nonlinearity sets in is indeed 
$k=k_{\mathrm{max}}$, and the Q-ball radius is approximately the wavelength of this mode at that time. Since Q-balls absorb nearly all the charge from the background field, their radius and charge are directly determined by the initial charge density in the Affleck-Dine field.

In the following, we will examine specific models and discuss their formation conditions.

\subsection{Q-ball formation in gravity-mediation model}

First, let us revisit the gravity-mediation model as studied in the context of the Affleck-Dine mechanism.
The effective potential in the gravity-mediation model is given by \eq{eq:gravitymed} or \begin{equation}
	V(\Phi) = m^{2}\left[ 1+K\log\frac{|\Phi|^{2}}{M_{G}^{2}} \right]|\Phi|^{2}
\end{equation}
where we take $v=0$ for simplicity. 
The condition for the growth of perturbations is:
\begin{equation}
	\frac{\vec{k}^{2}}{a^{2}} - \dot{\Omega}^{2} + m^{2}\left[ 1+3K+K\log\frac{|\Phi|^{2}}{M_{G}^{2}} \right] < 0
\end{equation}
When the field is coherently oscillating, the angular velocity in field space is equal to the effective mass such as 
\begin{equation}
\dot{\Omega}^{2} \simeq m_\text{eff}^{2} = m^{2}\left[ 1+K\log\frac{|\Phi|^{2}}{M_{G}^{2}} \right]
\end{equation}
Thus, the wavenumber range for growing modes (the growth band) is
\begin{equation}
0 \lesssim \frac{\vec{k}^{2}}{a^{2}} \lesssim 3m^{2}|K|
\end{equation}
This shows that Q-ball formation requires $K<0$. 
The most rapidly growing mode has a wavenumber given by
\begin{equation}
\frac{\vec{k}_{\mathrm{max}}^{2}}{a^{2}} = \frac{1}{16\dot{\Omega}^{2}}(7\dot{\Omega}^{2}+V'')(\dot{\Omega}^{2}-V'')
\simeq \frac{3}{2}m^{2}|K|
\label{eq:bestgrow}
\end{equation}
This mode is located near the center of the growth band and does not depend on the initial amplitude of the field.
The growth rate of this mode is
\begin{equation}
\alpha = \frac{3}{4}\frac{m^{2}|K|}{\dot{\Omega}}
\end{equation}

If the initial U(1) charge density is approximately $\mathcal{Q}\sim \omega|\Phi|^{2} \sim m|\Phi|^{2}$, 
then the typical charge of a Q-ball formed through this process is
\begin{equation}
	Q \sim \mathcal{Q}\times \left(\frac{\vec{k}_{\mathrm{max}}^{2}}{a^{2}}\right)^{-\frac{3}{2}} \sim |K|^{-\frac{3}{2}}\left(\frac{|\Phi|}{m}\right)^{2}
\label{eq:gracharge}
\end{equation}
Previous numerical simulations have found~\cite{Kasuya:2000wx}
\begin{equation}
	Q = \beta\left(\frac{|\Phi|}{m}\right)^{2} 
\end{equation}
where $\beta = \mathcal{O}(1)$ is a numerical constant. This is in good agreement with the analytic estimation.

\subsection{PQ-ball formation in the Peccei-Quinn model}
\label{PQballgene}

Now, we consider a Peccei-Quinn model with the potential given by \eq{eq:potential_PQ} or 
\begin{equation}
	V(\Phi) = m^{2}\left[ 1+k_{1}\log\frac{|\Phi|^{2}}{M^{2}} - k_{2}\log\frac{|\Phi|^{2}+v^{2}}{M^{2}} \right] |\Phi|^{2} ~~~~~~~~(k_{2}>k_{1}>0),
\end{equation}

If the field begins rotating in the angular direction with a large vacuum expectation value $|\Phi| \gg v$, then the shape of the potential at small $|\Phi|$ does not significantly affect the dynamics.
Thus, for perturbation growth, the discussion from the gravity-mediation model remains valid.
For small $v$, we can replace $K$ with $(k_{1}-k_{2})$ in the gravity-mediation model, leading to
\begin{equation}
\frac{\vec{k}_{\mathrm{max}}^{2}}{a^{2}} \simeq \frac{3}{2}m^{2}(k_{2}-k_{1})
\label{eq:grouthmode}
\end{equation}
The growth rate of this mode is
\begin{equation}
\alpha = \frac{3}{4}\frac{m^{2}(k_2-k_1)}{\dot{\Omega}}
\label{eq:grouthrate}
\end{equation}
The charge of a PQ-ball in this model is
\begin{equation}
	Q \sim (k_{2}-k_{1})^{-3/2}\left(\frac{|\Phi|}{m}\right)^{2}
\end{equation}
This result has essentially the same form as the charge in the gravity-mediation model, indicating that the formation mechanism in both cases follows the same principles.

\section{Absorbing boundary condition}
\label{sec:boundary}

To simulate PQ-ball decay, we consider a system where the stationary solution from Sec.~\ref{static_solution} exists in an otherwise vacuum region ($\Phi=0$). 
For this purpose, the boundary condition at infinity should be modified 
to mimic the effect of vacuum. This is achieved using the absorbing boundary condition (see, e.g., Ref.~\cite{Salmi:2012ta}), which prevents artificial reflections and allows outgoing waves to be absorbed, mimicking their escape to infinity. 

In the far region, where the field variation is small, the field equation can be approximated by the three-dimensional wave equation:
\begin{equation}
\frac{\partial^{2}\varphi}{\partial t^{2}} - \frac{\partial^{2}\varphi}{\partial r^{2}} - \frac{2}{r}\frac{\partial\varphi}{\partial r}
+m_{\varphi}^{2}\varphi = 0,
\label{eq:3Dwave}
\end{equation}
where $m_\varphi$ is the mass of the complex scalar field $\phi$. Using the Fourier transform:
\begin{equation}
\varphi(r,t) = \int d\omega\tilde{\varphi}(r,\omega)e^{i\omega t},
\end{equation}
this equation transforms into:
\begin{equation}
\frac{\partial^{2}\tilde{\varphi}}{\partial r^{2}} + \frac{2}{r}\frac{\partial\tilde{\varphi}}{\partial r} = (m_{\varphi}^{2}-\omega^{2})\tilde{\varphi}.
\end{equation}
For outgoing waves at large $r$, the solution is:
\begin{equation}
\tilde{\varphi}(r,\omega) = \frac{1}{r}\exp\left[ i\sqrt{\omega^{2}-m_{\varphi}^{2}}r \right]
\end{equation}
which satisfies:
\begin{equation}
\left. \left[ \frac{\partial}{\partial r}-i\sqrt{\omega^{2}-m_{\varphi}^{2}}+\frac{1}{r} \right]\tilde{\varphi} \right|_{r=r_{b}} = 0.
\end{equation}
After inverse Fourier transforming and expanding in $m/\omega$, this leads to the absorbing boundary condition:
\begin{equation}
\frac{\partial}{\partial t}\frac{\partial\varphi}{\partial r} + \frac{\partial^{2}\varphi}{\partial t^{2}}
+ \frac{1}{2}m_{\varphi}^{2}\varphi + \frac{1}{r}\frac{\partial\varphi}{\partial t} = 0.
\label{eq:abc}
\end{equation}
This ensures that waves exit without reflection, though low-energy components may still be partially reflected. The corresponding dimensionless form is:
\begin{equation}
	\frac{\partial}{\partial\tau}\frac{\partial\chi}{\partial\rho} + \frac{\partial^{2}\chi}{\partial\tau^{2}} + \frac{1}{\rho}\frac{\partial\chi}{\partial\tau} 
	+ \frac{1}{2}\left[ 1+k_{1}\left( 1 + \log|\chi_{\infty}|^{2} \right)-k_{2}\left( \frac{|\chi_{\infty}|^{2}}{|\chi_{\infty}|^{2} + \tilde{v}^{2}} + \log( |\chi_{\infty}|^{2} + \tilde{v}^{2} ) \right) \right]\chi = 0.
\end{equation}

When deriving the boundary condition (\ref{eq:abc}), we assumed that the mass term $m_{\varphi}$ in the wave equation (\ref{eq:3Dwave}) was a constant. However, in the equation we are currently considering (\ref{eq:EL2}), the coefficient of $\Phi$ is not constant. Nevertheless, as shown in Fig.~\ref{fig:gauss2}, the field $\Phi(r,t)$ approaches an approximately constant value $\Phi_{\infty}(t)$ at sufficiently large radii. Therefore, the coefficient in equation (\ref{eq:EL2}) can be treated as a constant (independent of $r$) and replaced by:
\begin{equation}
	m_{\varphi}^{2} \rightarrow m^{2}\left[ 1+k_{1}\left( 1 + \log\frac{|\Phi_{\infty}|^{2}}{M^{2}}\right)-k_{2}\left( \frac{|\Phi_{\infty}|^{2}}{|\Phi_{\infty}|^{2} + v^{2}} + \log\frac{|\Phi_{\infty}|^{2} + v^{2}}{M^{2}}\right) \right].
\end{equation}
This allows us to apply the absorbing boundary condition in regions where the spatial variation of the field is small.
Since $\chi_{\infty}=\Phi_{\infty}/M$ evolves over time, it must be updated at each time step during the computation.

\section{A concrete SUSY PQ model}
\label{sec:model}

In the main part of this paper, we assumed a potential of the form (\ref{eq:potential_PQ}) to guarantee both the finite VEV at the minimum and the existence of PQ-ball solution at large field value.
In SUSY, the running mass term with positive coefficient $(k_1>0)$ is obtained if $\Phi$ has Yukawa couplings to some other chiral superfields and negative coefficient $(k_2<0)$ may be obtained if it has gauge couplings~\cite{Enqvist:2000gq}.

Here we give an explicit example of the PQ model to realize (\ref{eq:potential_PQ}).\footnote{
    Non-SUSY version of the same model is found in Ref.~\cite{Sato:2018nqy}.
}
This model is an extension of the hadronic axion model~\cite{Kim:1979if,Shifman:1979if}.
We introduce a PQ superfield $\Phi=(\phi_1,\phi_2,\phi_3)$ that is a triplet under ``dark'' SU(2) gauge symmetry (SU(2)$_{\rm D}$) and chiral superfields $Q_1$ and $Q_2$ that are fundamental and anti-fundamental under SU(2)$_{\rm D}$.
The superpotetnial is given as follows:
\begin{align}
	W = y\,{\rm Tr}\left[Q_2 \Psi Q_1\right],
\end{align}
where $\Psi = \phi^a \sigma_a$.
Their charge assignments under the PQ symmetry (U(1)$_{\rm PQ}$), color SU(3)$_{\rm c}$ and SU(2)$_{\rm D}$ are summarized in Table~\ref{table:pq}.
The D-term potential is given by
\begin{align}
	V_D &= -\frac{g^2}{2} \left[ (\phi_2^\dagger\phi_3-\phi_2\phi_3^\dagger)^2
	+ (\phi_3^\dagger\phi_1-\phi_3\phi_1^\dagger)^2 + (\phi_1^\dagger\phi_2-\phi_1\phi_2^\dagger)^2 \right],\\
	&=2g^2 \left[\varphi_2^2\varphi_3^2 \sin^2\theta_{23} + \varphi_3^2\varphi_1^2 \sin^2\theta_{31} + \varphi_1^2\varphi_2^2 \sin^2\theta_{12} \right],
\end{align}
where $g$ is the SU(2)$_{\rm D}$ gauge coupling and
\begin{align}
	\phi_a = \varphi_a e^{i\theta_a},~~~~~~\theta_{ab} = \theta_a- \theta_b.
\end{align}
By using the SU(2)$_{\rm D}$ gauge degrees of freedom, one can choose $\varphi_2=\varphi_3=0$, for example.
Then the D-term potential vanishes and $\phi_1$ corresponds to the D-flat direction, which plays the role of $\Phi$ in the main text.
Then, by calculating the one-loop effective potential for the $\phi_1$ field, we obtain\footnote{
    To calculate it, one should figure out which degrees of freedom become massive when $\phi_1$ takes large field value.
    In particular, the gauge sector is a bit nontrivial.
    For bosonic part, two gauge bosons as well as $\varphi_2$ and $\varphi_3$ obtains mass of $\sim g|\phi_1|$.
    For fermionic part, two gauginos as well as fermionic partner of $\phi_2$ and $\phi_3$ become massive through Dirac-like mass terms. These two contributions cancel if there were no SUSY breaking masses.
}
\begin{align}
	V \simeq m^2|\phi_1|^2 \left( 1 + k_1 \log\frac{v_1^2+|\phi_1|^2}{\mu^2} - k_2 \log\frac{v_2^2+|\phi_1|^2}{\mu^2} \right),
    \label{V_concrete}
\end{align}
in the limit $|\phi_1|^2 \gg m_{\widetilde g}^2/g^2, m_{\widetilde Q}^2/y^2$,
where $m$ is the soft mass of $\Phi$, $\mu$ is the renormalization scale, $v_1^2\simeq m_{\widetilde Q}^2 / y^2$, $v_2^2 \simeq m_{\tilde g}^2/g^2$, and
\begin{align}
	k_1 \sim \frac{3}{16\pi^2}\frac{y^2m_{\widetilde Q}^2}{m^2},~~~~~~
	k_2 \sim \frac{3}{16\pi^2}\frac{g^2m_{\widetilde g}^2}{m^2}.
\end{align}
with $m_{\widetilde g}$ and $m_{\widetilde Q}$ being the SU(2)$_{\rm D}$ gaugino mass and soft scalar mass of $Q_1, Q_2$, respectively.
Thus, for a particular choice of parameters, this model can reproduce the desired form (\ref{eq:potential_PQ}).\footnote{
    Precisely speaking, the form (\ref{V_concrete}) is not valid for $|\phi_1| \lesssim v_1, v_2$, but the qualitative behavior of the potential does not change much and it is enough for our purpose.
}

\begin{table}
\begin{center}
\begin{tabular}{c|ccc}
\hline
~                           &$\Phi$ & $Q_1$ & $Q_2$    \\ \hline
U(1)$_{\rm PQ}$   & $+2$     & $-1$ & $-1$ \\ \hline
SU(3)$_{\rm c}$    & ${\bf 1}$  & ${\bf 3}$ & ${\bf \bar 3}$ \\ \hline
SU(2)$_{\rm D}$    & ${\bf 3}$ & ${\bf 2}$ &  ${\bf \bar 2}$  \\ \hline
\end{tabular}
\caption{Charge assingemnts on the fields.}
\label{table:pq}
\end{center}
\end{table}

\section{Kinetic misalignment mechanism}
\label{sec:kinetic}

Here we briefly review the idea of kinetic misalignment~\cite{Co:2019jts}.
The condition for kinetic misalignment to be effective is
\begin{equation}
\dot{\theta}_{i} = \dot{\theta}(T_{\text{osc}}) > 2m_{a}(T_{\text{osc}}),
\end{equation}
where $\theta$ is the angular component of $\Phi$, $m_a$ is its (temperature-dependent) mass and $T_{\rm osc}$ is the temperature at which the radial mode begins to oscillate.
At $T=T_{\text{osc}}$ the kinetic energy is still large enough for the field to cross the potential barriers, but as the universe expands, the kinetic energy decreases due to redshift, and at some later time, the field becomes trapped between the potential walls.\footnote{
    A similar axion dynamics in a model with multi axion-like particles has been considered in Refs.~\cite{Daido:2015bva,Daido:2015cba}.
} 
Denoting this temperature as $T=T'$ the trapping occurs when:
\begin{equation}
\dot{\theta}(T') = 2m_{a}(T').
\label{eq:kine_equal}
\end{equation}
Thus, the comoving axion number density is determined by the quantities at $T=T'$. Since the kinetic energy at this time is given by $f_{a}^{2}\dot{\theta}^{2}(T')/2$, we obtain 
\begin{equation}
n_{a}(T') = \frac{f_{a}^{2}\dot{\theta}^{2}(T')}{2m_{a}(T')} = Cf_{a}^{2}\dot{\theta}(T'),
\end{equation}
where $C$ is a constant that includes non-linear effects (deviations from harmonic oscillations) ignored in Eq.~(\ref{eq:kine_equal}). It is estimated as $C\simeq2$~\cite{Co:2019jts}. The U(1) charge density associated with PQ symmetry, $n_\theta$,  redshifts as $n_{\theta}\propto a^{-3}$, with $a$ being the cosmic scale factor, due to charge conservation.
Therefore, the quantity
\begin{equation}
	Y_{\theta}\equiv \frac{n_{\theta}}{s} = \frac{f_{a}^{2}\dot{\theta}}{s},
\end{equation}
remains invariant under cosmic expansion until $T=T'$.
At $T=T'$, $n_\theta$ and $n_a$ are roughly equal and $n_a \propto a^{-3}$ afterwards.
Therefore, the axion energy-to-entropy ratio is
\begin{equation}
\frac{\rho_{a}}{s} = m_{a}\frac{n_{a}(T')}{s(T')} = Cm_{a}Y_{\theta},
\label{eq:kin_energy}
\end{equation}
and the axion density parameter is given by
\begin{equation}
\Omega_{a}h^{2} \simeq \Omega_\text{DM}h^{2}\left( \frac{10^{9}\text{GeV}}{f_{a}} \right) \left(\frac{Y_{\theta}}{40} \right),
\end{equation}
where $\Omega_\text{DM}h^{2} \simeq 0.12$ is the density parameter of dark matter~\cite{Planck:2018vyg}. 
To determine $Y_{\theta}$, a detailed analysis of the scalar dynamics around
$T=T_\text{osc}$ in a concrete model is necessary.

\bibliography{reference}

\end{document}